\documentstyle{jsc}
\title[Computation of Conserved Densities]
{{Symbolic Computation of Conserved Densities for Systems of Nonlinear
Evolution Equations}
\footnote{Research supported in part by 
the National Science Foundation under Grants CCR-9300978 and CCR-9625421.}}
\author[\"{U}. G\"{o}kta\c{s} and W. Hereman]
{\"{U}NAL G\"{O}KTA\c{S} AND WILLY HEREMAN \\
Department of Mathematical and Computer Sciences, \\
Colorado School of Mines, Golden, CO 80401-1887, U.S.A.} 
\date{26 February 1997}

\pagerange{\pageref{firstpage}--\pageref{lastpage}}
\def\LaTeX{L\kern-.25em\raise.425ex\hbox{a}\kern-.075em\TeX}

\newtheorem{theorem}{Theorem}
\newtheorem{example}{Example}

\renewcommand{\theenumi}{\arabic{enumi})}

\begin{document}
\label{firstpage}
\maketitle
\begin{abstract}
A new algorithm for the symbolic computation of polynomial conserved
densities for systems of nonlinear evolution equations is presented.
The algorithm is implemented in {\it Mathematica}. 
The program {\bf condens.m} automatically carries out the lengthy         
symbolic computations for the construction of conserved densities.
The code is tested on several well-known partial differential equations 
from soliton theory.
For systems with parameters, {\bf condens.m} can be used to determine
the conditions on these parameters so that a sequence of conserved
densities might exist.
The existence of a large number of conservation laws is a predictor for 
integrability of the system.
\end{abstract}

\section{Introduction}
An indication that certain evolution equations might have remarkable
mathematical properties came with the discovery of an infinite number 
of conservation laws for the Korteweg-de Vries (KdV) equation,
$u_t + u u_x + u_{3x} = 0 $.
The conserved quantities $u$ and $u^2$, corresponding to conservation of 
momentum and energy, respectively, were long known, and Whitham (1974) 
had found a third one, $u^3 - 3 u_x^2,$ which corresponds to Boussinesq's 
moment of instability. 
Zabusky and Kruskal found a fourth and fifth. However, the search for 
additional conserved densities was halted due to a mistake in 
their computations. 
Miura eventually continued the search and, beyond the missing sixth, 
found an additional three conserved densities (Newell, 1983). 
It became clear that the KdV equation had an infinite sequence of 
conservation laws, later proven to be true. 

The existence of an infinity of conserved densities was an important link 
in the discovery of other special properties of the KdV equation 
(Zakharov, 1990). 
It led, for example, to the construction of the Miura transformation, 
which connects the solutions of the KdV and modified KdV equations. 
Consequently, the famous Lax pair was found, which associates a couple of 
linear equations to the KdV equation.  
From that, the inverse scattering technique (IST) for direct linearization of 
integrable equations was developed, and it was then shown that the KdV 
equation, and many other integrable equations, admit bi-Hamiltonian structures.

There are several motives to construct conserved densities of
partial differential equations (PDEs) explicitly.
The first few conservation laws have a physical interpretation.
Additional ones may facilitate the study of both quantitative and 
qualitative properties of solutions (Scott {\em et al.\/}, 1973). 
Furthermore, the existence of a sequence of conserved densities
(perhaps with gaps) predicts integrability. 
Yet, the non-existence of conserved quantities does not preclude integrability.
There are indeed equations, viz. dissipative equations, with only 
{\em one} conserved density, which can be directly integrated. 
The most notable is the Burgers equation, which can be transformed
into the {\it linear} heat equation via the Cole-Hopf transformation
(Zakharov, 1990).

Yet another compelling argument to explicitly construct conserved 
densities relates to the numerical solution of PDEs. 
It is desirable that the semi-discretization conserves the discrete 
analogues of the continuous conserved quantities.
In particular, the conservation of a positive definite quadratic quantity 
may prevent the occurrence of nonlinear instabilities in the numerical 
scheme. 
The use of conservation laws in solving the Boussinesq equation 
numerically has been illustrated in Hickernell (1983).
Sanz-Serna (1982) describes a scheme for the integration in time of PDEs, 
which is explicit and capable of conserving discretized quadratic functionals. 
Since $u \;{\rm and}\; u^2$ are conserved densities for the KdV equation, 
a discrete scheme should have conservation of 
momentum, $\sum_{j} U_{j}^{n},$ and energy $\sum_{j} {[U_{j}^{n}]}^2.$ 
In Sanz-Serna (1982), an explicit self-adaptive conservative scheme with
conservation of energy
and  momentum is given. 
Conservation of energy implies boundedness of the solutions, and therefore 
obviates the occurrence of any blowup phenomena.
For more details about numerical applications of conservation laws see 
LeVeque (1992). 

We present a new symbolic algorithm for computing closed-form 
polynomial-type conservation laws for systems of nonlinear 
evolution equations. Our algorithm is also applicable to wave equations,
viz. the Boussinesq equation, provided that the PDE (of higher-order in time)
can be written as a system of evolution equations (first order in time).
In contrast to fairly complicated algorithms designed by Bocharov (1991), 
Gerdt and Zharkov (1990), and Gerdt (1993), we introduce an algorithm that 
is based in part on ideas presented in 
Hereman and Zhuang (1995), Ito (1994), and Ito and Kako (1985), 
Kruskal {\em et al.\/} (1970), Kodama (1985),
Miura {\em et al.\/} (1968), Verheest and Hereman (1995), and 
Willox {\em et al.\/} (1995). 
Our algorithm has the advantage that it is fairly straightforward to 
implement in any symbolic language. 
We also present a software package {\bf condens.m}, written in 
{\it Mathematica} syntax, which automates the tedious computation
of closed-form expressions for conserved densities and fluxes. 

In Section 2, we give the definitions of conservation law, density and flux. 
We also state a theorem from calculus of variations about the Euler-Lagrange 
equations, which will play a role in our algorithm. 
The remainder of the section is devoted to a detailed exposition of the
algorithm, which was implemented in {\it Mathematica}, and successfully 
tested on many well-known evolution systems from soliton theory. 

In Section 3, we address applications of our program {\bf condens.m}. 
In particular, we show how it could be used in a search for integrable 
fifth-order evolution equations of KdV type, where we retrieved all the 
known integrable cases. 
We carried out a similar computer search for a parameterized class of
seventh-order evolution equations. 
More examples and test results are presented in Section 4.
In Section 5, we describe the usage of our code, and indicate 
its capabilities and limitations.
In Section 6, a comparison with other programs is given. 
Also, several ongoing projects are briefly addressed.

\section{Computation of Conserved Densities}

\subsection{Definitions}
For simplicity, consider a single PDE, 
\begin{equation}\label{pde}
\Delta (x,t,u(x,t)) = 0,
\end{equation}
where $t \in I\!\!R $ denotes time, $x \in I\!\!R$ is the spatial 
variable, and $u(x,t) \in I\!\!R$ is the dependent variable.
A {\em conservation law} is of the form
\begin{equation}\label{conslaw}
{\rm D}_{t} \rho + {\rm D}_{x} J = 0, 
\end{equation}
which is satisfied for all solutions of (\ref{pde}).
The functional $\rho(x,t)$ is the {\em conserved density\/}, 
$J(x,t)$ is the associated {\em flux\/}.
Both are, in general, functions of $x, t, u,$ and its partial derivatives 
with respect to $x.$
Furthermore, ${\rm D}_{t}$ denotes the total derivative with respect to t; 
${\rm D}_{x}$ the total derivative with respect to $x$ 
(Ablowitz and Clarkson, 1991).
Specifically, $\rho$ is a {\it local} conserved density if $\rho$ is a 
local functional of $u$, i.e. if the value of $\rho$ at any $x$ depends 
only on the values of $u$ in an arbitrary small neighborhood of $x.$ 
If $J$ is also local, then (\ref{conslaw}) is a {\it local conservation law}.
In particular, if $\rho$ is a polynomial in $u$ and its $x$ derivatives,
and does not depend explicitly on $x$ or $t$, then $\rho$ is called
a {\it polynomial} conserved density.
If $J$ is also such a polynomial, then (\ref{conslaw}) is
called a {\it polynomial conservation law}. 
There is a close relationship between constants of motion and 
conservation laws.
Indeed, for polynomial-type $\rho$ and $J$, integration of 
(\ref{conslaw}) yields
\begin{equation} \label{const}
P = \int_{-\infty}^{+\infty} \rho \; dx = {\rm constant},
\end{equation}
provided that $J$ vanishes at infinity.
For ordinary differential equations, the $P\/$'s are constants of motion.
\begin{example}
{\rm The most famous evolution equation from soliton theory, the 
KdV equation (Miura, 1968),
\begin{equation} \label{KdVeq} 
u_t + u u_x + u_{3x} = 0 , 
\end{equation} 
is known to have infinitely many polynomial conservation laws. 
The first three are 
\begin{eqnarray}
& &(u)_t  + \left ( {1 \over 2} u^2 + u_{2x} \right )_x = 0,\\ 
& & \left (u^2 \right)_t +  
   \left ( {2 \over 3} u^3 - {u_x}^2 + 2 u u_{2x} \right)_x = 0,\\ 
& &\left ( u^3 - 3 {u_x}^2 \right )_t  + 
   \left ( {3 \over 4} u^4 - 6 u {u_x}^2 + 3 u^2 u_{2x} + 3 {u_{2x}}^2 -
           6 u_x u_{3x} \right)_x = 0.
\end{eqnarray}
}
\end{example}
The first two express conservation of momentum and energy, respectively.
They are easy to compute by hand. 
The third one, less obvious and requiring more work, corresponds
to Boussinesq's moment of instability (Newell, 1983).

Observe that the KdV equation and its densities $\rho = u, u^2 $ and 
$u^3 - 3 {u_x}^2 $ are all invariant under the scaling symmetry
\[
(x,t,u) \rightarrow (\lambda x, \lambda^3 t, \lambda^{-2} u), 
\]
where $\lambda$ is a parameter.
Stated differently, $u$ carries the weight of two derivatives with respect 
to $x;$ denoted symbolically by 
\[ u \sim {\partial^2 \over \partial x^2}.
\]

Scaling invariance, which is a special Lie-point symmetry, is an intrinsic 
property of many integrable nonlinear PDEs.
Our algorithm exploits this idea in the construction of conserved densities.

\subsection{Euler Operator}

We introduce a tool from calculus of variations: the Euler operator, which is
very useful for testing if an expression is a total derivative 
(Ito and Kako, 1985; Olver, 1986), without having to carry out any 
integrations by parts.
\begin{theorem}\label{thm5}
If $f = f(x,y_1,\ldots,y_1^{(n)},\dots,y_{N},\ldots,y_{N}^{(n)})$,
then ${\cal L\/}_{\vec{y}}(f) \equiv \vec{0},$
if and only if 
\vskip 3pt
\noindent
${\displaystyle f= \frac{\mbox{d}}{dx} g}$,
where $  g = g(x,y_1,\ldots,y_1^{(n-1)},\dots,
   y_N,\ldots,y_N^{(n-1)}) $.
\end{theorem}
\noindent
In this theorem, for which a proof can be found in
Olver (1986, pp. 252),
$\vec{y} = [y_1,\ldots,y_N]^{T},$
\vskip 4pt
\noindent
$ {\cal L\/}_{\vec{y}}(f) = [{\cal L\/}_{y_1}(f),\ldots,
{\cal L\/}_{y_N}(f)]^T, \; $
$ \vec{0} = [0,\ldots,0]^{T}, $ with $T$ for transpose, and where  
\[{\cal L\/}_{y_i} = \frac{\partial }{\partial{y_i}} -
\frac{d}{dx}( \frac{\partial }{\partial{{y_i}'}})+
\frac{d^2}{dx^2}( \frac{\partial }{\partial{{y_i}''}})+\cdots+
(-1)^n \frac{d^n}{dx^n} (\frac{\partial }{\partial{{y_i}^{(n)}}}),
\]
is the {\em Euler operator} (or variational derivative). 
We will use this theorem in our algorithm. 

\subsection{Algorithm}

We now describe our algorithm to compute polynomial conservation laws
(\ref{conslaw}) for systems of nonlinear evolution equations of
order $n,$
\begin{equation}\label{multisys}
{\vec u}_t = 
{\vec {\cal F}} ({\vec u}(x,t),{\vec u~}'(x,t),...,{\vec u}^{(n)}(x,t)), 
\end{equation}
where ${\vec u} = [u_1,\ldots,u_N]^T\/$, or, component-wise,
\begin{equation} \label{givensystem}
u_{i,t} + {\cal F}_i (u_j, u_{j}^{'}, u_{j}^{''}, \ldots, u_{j}^{(n)} ) = 0,
\quad i=1,2,\ldots,N,\quad j=1,2,\ldots,N,
\end{equation}
where 
\[{\displaystyle u_{i,t} \buildrel \rm def \over =
       \frac{\partial{u_i}}{\partial{t}} ,\; \quad \;
       {u_j}^{(n)} = u_{j,nx} \buildrel \rm def \over =
       \frac{\partial^n{(u_j)}}{\partial{x^n}},} 
\]
and all components of ${\vec u} $ depend on $x$ and $t.$ 
\vskip 4pt
\noindent
Our goal is to compute the densities
$\rho({\vec u},...,{\vec u}^{(m_1)}) $ and the fluxes
$ J({\vec u},...,{\vec u}^{(m_2)}) $ of order $m_1$ and $m_2$, respectively.
There will be no restriction on the order $n$ of the system (\ref{multisys}),
or on its degree of nonlinearity. But, all the components of ${\vec {\cal F}}$ 
must be polynomial in their arguments.
Furthermore, we only consider systems of evolution equations in $t$ with
one spatial variable $x$.

In our algorithm, we tacitly assume that we have an evolution equation 
for every dependent variable $u_i$.
In cases where there are more dependent variables than equations, 
one can always add trivial evolution equations $u_{i,t} = 0.$
Further on, we use the notation $u_{j,nx}$ instead of ${u_j}^{(n)}$ because 
it is closer to the notation in our code. 
\vfill
\newpage
\noindent
{\bf Step 1$\;\;\;$Determine the weights of variables and parameters}
\vskip 4pt
We define the {\em weight} of a variable as the number of partial
derivatives with respect to $x$ the variable carries, and the 
{\em rank} of a term as the total weight in terms of partial 
derivatives with respect to $x$ (Kruskal {\em et al.\/}, 1970; 
Miura {\em et al.\/}, 1968). 
The rank is assumed to be nonnegative and rational.

For the system (\ref{givensystem}), we first try to determine 
the weights (scaling properties) of all the variables. 
We assume that all terms in a particular equation have the same rank. 
We call this property {\it uniformity in rank}. 
Different equations in (\ref{givensystem}) can have different ranks. 
Having defined the weight of a variable in terms of $x-$derivatives,
we (can) pick
$ w(\frac{\partial{}}{\partial{x}}) = 1,\ldots,
  w(\frac{\partial^n{}}{\partial{x^n}}) = n, $ 
where $w\/$ returns the weight of its argument.

For simple systems, in particular those with uniform rank, 
only the variables $u_i$ and $\frac{\partial{}}{\partial{t}}$ have weights. 
However, to be able to handle more general systems, we allow for 
constant parameters to be introduced and also for these 
parameters to carry weights. 
The trick of introducing parameters with weights allows one to handle 
equations without uniform rank. 
Let us assume that there are $P$ such parameters in the system, 
denoted by $p_i, \; i=1,2,...,P.$
Thus, the extended list of {\em variables} that carry weights is
$\{\frac{\partial{}}{\partial{t}},u_1,u_2,\ldots,u_N,p_1,p_2,\ldots,p_P\}$. 

Weights must be nonnegative and rational, except for 
$w(\frac{\partial}{\partial t}),$ which may be positive, zero, or negative.  
More precisely, the weight of at least one $u_i$ must be positive; 
the weights of the parameters $p_i$ must be nonnegative. 
We proceed with the following steps: 
\begin{enumerate}
\item[(a)] Take the $i^{th}$ equation in (\ref{givensystem}). 
Denote the number of terms in the equation by $K_i.$ 
\item[(b)] Compute the rank $r_{i,k}\/$ of the $k^{th}$ term in 
the $i^{th}$ equation as follows: 
\newline
\noindent
$ 
\displaystyle{ r_{i,k} = d(x) + d(t) \; w(\frac{\partial{}}{\partial{t}}) + 
            \sum_{j=1}^{N} g(u_j) \; w(u_j)+ \sum_{j=1}^{P} 
            g(p_j) \; w(p_j),\quad k=1,2,\ldots,K_i, }
$
\newline
\noindent
where $g\/$ returns the degree of nonlinearity of its argument,
$d\/$ returns the number of partial derivatives with respect to its 
argument. For evolution equations $d(t)$ is either zero or one. 
\item[(c)] Assuming uniformity in rank in the $i^{th}$ equation, 
form the linear system
\[ A_i = \{r_{i,1} = r_{i,2} = \cdots = r_{i,K_i} \}. \] 
\item[(d)] Repeat steps (a) through (c) for all of the equations 
in (\ref{givensystem}).
\item[(e)] Gather the equations $A_i$ to form the global linear system
$\displaystyle {\cal A} = \bigcup_{i=1}^{N} A_i. $ 
\item[(f)] Solve $\cal A$ for the $N+P+1$ unknowns $w(u_j)$, $w(p_j)$ 
and $w(\frac{\partial{}}{\partial{t}})$.
\end{enumerate}
If the solution of the system $\cal A$ still has free, consider two cases:
\begin{enumerate}
\item If two or more weights are undetermined, prompt the user to enter 
choices.
\item If only one weight is free, say $w(u_k),$ take the equations obtained
in (f), set their left hand sides equal to one, and solve them piece by 
piece for $w(u_k).$ Include the choice $w(u_k) = 1$.
For all choices for $w(u_k)$ test:
(i) if $w(u_k)$ is negative, increment it until it is positive;
(ii) reject $w(u_k)$ if any other weight is negative.
Continue with the smallest integer value for $w(u_k)$ if present, else
take the smallest fractional value. 
This produces at most {\it one} positive value for the free weight, out of 
possibly infinitely many choices. 
If the algorithm fails, prompt the user to enter a value for $w(u_k).$ 
\end{enumerate}
\noindent
{\bf Step 2$\;\;\;$Construct the form of the density}
\vskip 4pt
The second step involves the construction of the polynomial 
density with a prescribed rank $R.$
All the terms in the density $\rho$ must have that same rank $R$. 
Since we may introduce parameters with weights, the fact that 
the density will be a sum of monomials of uniform rank does not 
necessarily imply that the density must be uniform in rank with respect to 
the dependent variables. 
Note that the rank $R$ can differ from any of the ranks of the 
equations in (\ref{givensystem}). The rank $R$ must be positive and
rational. 

Let ${\cal V}=\{v_1,v_2,\ldots,v_Q \}$ be the sorted list of all the 
variables with positive weights, including the parameters $p_i,$
but excluding $\frac{\partial{}}{\partial{t}}.$ 
The variables are ordered according to descending weights: $w(v_1)$ is
the largest weight, $w(v_Q)$ is the smallest. 
The following procedure is used to determine the form of the density 
of rank $R:$
\begin{enumerate}
\item[(a)] Form all monomials of rank $R$ or less by taking combinations 
of the variables in $\cal V.$ 
Recursively, form sets consisting of ordered pairs. In each pair,
the first component has a specific combination of different powers of the 
variables, the second component has the weight of the first component.
\vskip 2pt
\noindent
Set ${\cal B}_{0} = \{(1;0)\}$ and proceed as follows:
\vskip 2pt
\noindent
{\bf For} $q=1$ through $Q$ {\bf do} 
\begin{itemize}
\item[] {\bf For} $m=0$ through $M-1$ {\bf do}
\begin{itemize}
\item[] Form $\displaystyle B_{q,m} = \bigcup_{s=0}^{b_{q,m}}
    \{ (T_{q,s}; W_{q,s}) \}, \; $
where $M$ is the number of pairs in ${\cal B}_{q-1}, \;$ 
\vskip 2pt
\noindent
$ T_{q,s} = T_{q-1,m} \; {v_q}^s, \;$ 
$ W_{q,s} = W_{q-1,m} + s \; w(v_q), \;$ 
$(T_{q-1,m};W_{q-1,m})$ is the 
\vskip 3pt
\noindent
${(m+1)}^{st}$ ordered pair 
in ${\cal B}_{q-1}, \;$ and $ b_{q,m} = \lbrack\!\lbrack
 \frac{R- W_{q-1,m}}{w(v_q)} \rbrack\!\rbrack $ 
\vskip 3pt
\noindent
is the maximum allowable power of $v_q.$
\end{itemize}
\item[] Set $\displaystyle {\cal B}_{q} = \bigcup_{m=0}^{M-1} B_{q,m}.$
\end{itemize}
\item[(b)] Set ${\cal G} = {\cal B}_{Q}.$ 
Note that ${\cal G}$ has all possible combinations of powers of the 
variables that produce rank $R$ or less.  
\vskip 1pt
\noindent
\item[(c)] Introduce partial derivatives with respect to $x$. 
For each pair $( T_{Q,s}; W_{Q,s})$ in ${\cal G}$, apply
$\displaystyle \frac{\partial^{\ell}{}}{\partial{x^{\ell}}}$ 
to the term $T_{Q,s}$, provided $\displaystyle \ell=R- W_{Q,s} $ is integer. 
This introduces just enough partial derivatives with respect to $x$, so that 
all the pairs retain weight $R.$ 
Gather in a set ${\cal H}$ all the terms that result from computing the 
various $\displaystyle 
\frac{\partial^{\ell}{(T_{Q,s})}}{{\partial{x^{\ell}}}}.$ 
\item[(d)] Remove those terms from ${\cal H}$ that can be written as a total
derivative with respect to $x$, or as a derivative up to terms
kept prior in the set. Call the resulting set ${\cal I},$ which consists of 
the {\em building blocks\/} of the density $\rho$ with desired rank $R.$ 
\vskip 1pt
\noindent
\item[(e)] If ${\cal I}$ has $I$ elements, then their linear combination 
will produce the polynomial density of rank $R.$  Therefore, 
\[ \displaystyle{
\rho = \sum_{i=1}^{I} c_i \; {\cal I}(i), 
} 
\]
where ${\cal I}(i)$ denotes the $i^{th}$ element in ${\cal I},$ and
$c_i$ are numerical coefficients, still to be determined. 
\end{enumerate}
\vfill
\newpage
\noindent
{\bf Step 3$\;\;\;$Determine the unknown coefficients in the density}
\vskip 6pt
Recall that a conservation law is of the form 
$\displaystyle{ D_t \rho + D_x J = 0, }$ or 
$ \displaystyle{ D_t \rho = -D_x J, }$ 
which means that $D_t \rho$ must be the negative of the 
$x-$derivative of a functional $J.$

After computation of $D_t \rho,$ remove all ${(u_{i,t})}^{(j)},\; 
j=0,1,2,...$ from the expression, using the evolution equations in 
(\ref{givensystem}). 
The resulting expression for $D_t \rho$ must be a total $x-$derivative
of some expression.
To verify this we use Theorem \ref{thm5} and apply the Euler operator.
We require that the resulting Euler-Lagrange equations vanish identically
by the appropriate choice of the coefficients $c_i.$ 
This leads to a {\it linear} system for the ${c_i}.$ 
The system must be analyzed and solved for the unknown $c_i.$
In general, the procedure is as follows:
\begin{enumerate}
\item[(a)] Compute $D_t \rho$ and replace all ${(u_{i,t})}^{(j)},
i=0,1,...,N$ and $j=0,1,2,...$ using the evolution equations in 
(\ref{givensystem}). 
\vskip 4pt
\noindent
\item[(b)] The resulting expression, called $E$, must equal $D_x (-J)$ 
for some functional $J.$ 
\noindent
Apply to $E$ the Euler operator ${\cal L}$ from Theorem \ref{thm5}.  
If $E$ is completely integrable no terms will be left, 
i.e. $ {\cal L\/}_{\vec{u}}(E) \equiv {\vec 0}$. 
If terms remain, set them equal to zero and form the linear 
system for the ${c_i}$, denoted by  ${\cal S}.$ 
\vskip 4pt
\noindent
\item[(c)] Depending on whether or not there are parameters in 
${\cal S},$ two cases occur:              
\begin{itemize}
\item[(i)] If the only unknowns in ${\cal S}$ are $c_i$, solve
${\cal S}$ for the $c_i$. 
Substitute the (non-empty) solution into the expression of $\rho$ to obtain
its final form. 
\vskip 4pt
\noindent
\item[(ii)] If in addition to the coefficients $c_i$ there are parameters 
$p_i$ in ${\cal S},$ then determine the conditions on these parameters, 
so that a density in the given rank exists for at least some $c_i$ nonzero.
These {\em compatibility conditions} assure that the system ${\cal S}$ 
has non-trivial solutions.
Obviously, all $c_i$ equal to zero would give the trivial (zero) density.
Solving the compatibility conditions may lead to different densities of 
the same rank, corresponding to different choices of the parameters. 
Thus, generating the compatibility conditions enables one to filter out all 
the cases for which there exists a nontrivial density of given rank. 
Let ${\cal C}$ $=$ $\{ c_1,c_2,\ldots,c_I \}$ be the set of all the 
coefficients that appear in the density. 
In order to determine all possible compatibility conditions, 
proceed as follows:
\vskip 4pt
\noindent
Under the assumption that no $p_i$ in $\cal S$ is {\it zero\/},
analyze the system $\cal S.$ First determine the coefficients $c_i$ 
that always must be zero. Exclude these $c_i$ from $\cal C$
and set $i = i'$, where $i'$ is the smallest index of the $c_j$ that
remain in $\cal C$.
\vskip 4pt
\noindent
{\bf While} ${\cal C} \neq \{ \}$ {\bf do}:
\begin{itemize}
\item[] For the building block ${\cal I}(i)$ with coefficient $c_i$ 
to appear in $\rho$, one needs $c_i \not= 0 .$ Therefore, set $c_i = 1$ and
eliminate all the other $c_j$ from ${\cal S}.$ 
This gives compatibility conditions consistent with the presence of 
the term $c_i {\cal I}(i)$ in $\rho.$ 
\vskip 3pt
\noindent
{\bf If} $\cal S$ becomes inconsistent, or compatibility conditions 
require some of the parameters to be zero, 
\vskip 5pt
\noindent
{\bf then}:
\begin{itemize}
\item[] $c_i$ must be zero. 
Hence, set 
$ {\cal C} = {\cal C} \backslash \{c_i\},$ and $i=i',$ where $i'$ is the 
smallest index of the ${c_j}$ that remain in ${\cal C},$
\end{itemize}
{\bf else}:
\begin{itemize}
\item[] Analyze the compatibility conditions and for each resulting branch;
solve the system ${\cal S}$ for $c_j$, and substitute the solution into 
the expression of $\rho$ to obtain its final form. 
Then, collect those ${c_j}$ which are zero for {\it all} of the branches 
into a set ${\cal Z}.$ 
Since the $c_i$ in ${\cal Z}$ might not have occurred in any of the
densities yet, set 
$ {\cal C} = {\cal C} \cap {\cal Z},$ and $i=i',$ where
$i'$ is the smallest index of the $c_j$ that are still in ${\cal C}.$ 
\end{itemize}
\end{itemize}
\end{itemize}
\item[(d)] Compute $ J = - \int E \; dx, $ via integration by parts.
\end{enumerate}
The three examples below illustrate the algorithm. In the first example, 
we show Steps 1 and 2 in detail. The second and third example illustrate
details of Step 3. The examples 4.1 and 4.2 illustrate what happens if 
there are free weights.
\vskip 3pt
\noindent
\begin{example}
{\rm
The wave equation, 
\begin{equation} \label{Bouseq}
u_{tt} - u_{2x} + 3 u u_{2x} + 3 {u_x}^2 + \alpha u_{4x} = 0,
\end{equation}
was proposed by Boussinesq to describe surface water waves whose horizontal
scale is much larger than the depth of the water 
(Ablowitz and Clarkson, 1991; Hickernell, 1983).
Conservation laws play a key role in the study of (\ref{Bouseq})
for they can be used to prove that solutions are bounded for certain sets 
of initial conditions 
(Hickernell, 1983),
or, conversely, to prove that solutions fail to exist after a finite time.

For computing conservation laws, we rewrite (\ref{Bouseq}) as a system of 
first-order equations,
\begin{eqnarray} \label{Boussys1}
u_t + v_x &=& 0, \nonumber \\
v_t + u_x - 3 u u_x - \alpha u_{3x} &=& 0,
\end{eqnarray}
where $v$ is an auxiliary dependent variable.
It is easy to verify that the terms $u_x$ and $\alpha u_{3x}$ in the second 
equation do not allow for uniformity in rank. 
To circumvent the problem we use a trick: we introduce an 
auxiliary parameter $\beta$ with 
(unknown) weight, and replace (\ref{Boussys1}) by
\begin{eqnarray} \label{Boussys2}
u_t + v_x &=& 0,\nonumber \\
v_t + \beta u_x - 3 u u_x - \alpha u_{3x} &=& 0,
\end{eqnarray}
or, in our notation
\begin{eqnarray} \label{prgboussys1}
u_{1,t} + u_{2,x} &=& 0, \nonumber \\
u_{2,t} + \beta \; u_{1,x} - 3 u_1 u_{1,x} - \alpha \; u_{1,3x} &=& 0.
\end{eqnarray}
Using the procedure in Step 1, we obtain the weights by first computing
\[
\begin{array}{rclcrcl}
r_{1,1} &=& 1 \; w(\frac{\partial{}}{\partial{t}}) + 1 
\; w(u_1),& \quad\quad & 
r_{1,2} &=& 1+ 1 \; w(u_2), \\[.6mm]
r_{2,1} &=& 1 \; w(\frac{\partial{}}{\partial{t}}) + 1 
\; w(u_2),& \quad\quad & 
r_{2,2} &=& 1+ 1 \; w(u_1) + 1 \; w(\beta), \\[.6mm]
r_{2,3} &=& 1+ 2 w(u_1),& \quad\quad & 
r_{2,4} &=& 3+ 1 \; w(u_1), 
\end{array}
\]
then forming the systems
\[
\begin{array}{rclcrclcrcl}
A_1 &=& \{ r_{1,1} = r_{1,2} \},&\!\!\!\!\!& 
A_2 &=& \{ r_{2,1} = r_{2,2} = r_{2,3}=r_{2,4} \}, &\!\!\!\!\!\!& 
\; {\rm and} \;\;\; {\cal A} &=& A_1 \cup A_2, 
\end{array}
\]
and, finally, solving 
$\cal A$ for $w(u_1),w(u_2),w(\frac{\partial{}}{\partial{t}})$ and 
$w(\beta).$  This yields
\[ w(u_1) = 2, \; w(u_2) = 3, \; w(\beta) = 2, \;\; {\rm and} \;
   w(\frac{\partial{}}{\partial{t}}) = 2. \]
Hence, the scaling properties of (\ref{prgboussys1}) are such that
\[
 u_1 \sim \beta \sim {\partial^2 \over \partial x^2}, \quad
 u_2 \sim {\partial^3 \over \partial x^3}, \quad
 \frac{\partial{}}{\partial{t}} \sim \frac{\partial^2{}}{\partial{x^2}},
\]
which expresses that (\ref{prgboussys1}) is invariant under the 
scaling symmetry
\[
(x, t, u_1, u_2, \beta) 
\rightarrow 
(\lambda x, \lambda^2 t, \lambda^{-2} u_1, 
\lambda^{-3} u_2, \lambda^{-2} \beta). 
\]
Let us construct the form of the density with rank $R=6$.
Here, ${\cal V} = \{u_2, u_1, \beta \},$ hence, $v_1 = u_2, v_2 = u_1 $ and
$v_3 = \beta$ and, obviously, $Q = 3.$
We follow the procedure outlined in 
Step 2:
\begin{enumerate}
\item[(a)] For $\boldmath q=1,m=0$:
\vskip 4pt
\noindent
$ b_{1,0} = \lbrack\!\lbrack \frac{6}{3} \rbrack\!\rbrack = 2. $ 
Thus, with 
$ T_{1,s} = {u_2}^s, \; {\rm and} \; W_{1,s} = 3 s, \; $ where $s=0,1,2,$ 
we obtain
\[ {\cal B}_{1} = B_{1,0} = \{ (1;0),(u_2;3),({u_2}^2;6) \} . \]
\vskip 6pt
\noindent
For $\boldmath q=2,m=0$:
\vskip 4pt
\noindent
$ b_{2,0} = \lbrack\!\lbrack \frac{6-0}{2} \rbrack\!\rbrack =3.$
So, with 
$ T_{2,s} = {u_1}^s, \; {\rm and} \; W_{2,s} = 2 s, $ with $ s = 0,1,2,3,$
we obtain 
\[ B_{2,0} = \{ (1;0),(u_1;2),({u_1}^2;4),({u_1}^3;6) \}.\]
\vskip 6pt
\noindent
For $\boldmath q=2,m=1$:
\vskip 4pt
\noindent
we obtain
\[ B_{2,1} = \{ (u_2;3), (u_1 u_2;5) \}, \] 
since
$ b_{2,1} = \lbrack\!\lbrack \frac{6-3}{2} \rbrack\!\rbrack =1, \;$
$ T_{2,s} = 
u_2 \; {u_1}^s, \; {\rm and} \; W_{2,s} = 3 + 2 s, \;\;{\rm and}\;\; s=0,1. $ 
\vskip 10pt
\noindent
For $\boldmath q=2,m=2$:
\vskip 3pt
\noindent
$ b_{2,2} = \lbrack\!\lbrack \frac{6-6}{2} \rbrack\!\rbrack =0.$
Therefore,
$\displaystyle B_{2,2} = \{ ({u_2}^2;6) \}.$
Hence,
\[ {\cal B}_{2} = 
   \{ (1;0),(u_1;2),({u_1}^2;4),({u_1}^3;6),(u_2;3), (u_1 u_2;5),
   ({u_2}^2;6) \}. \] 
For $\boldmath q=3$: 
\vskip 3pt
\noindent
we introduce possible powers of $\beta.$ So, 
\[
\begin{array}{lclclcl}
B_{3,0} &=& \{ (1;0),(\beta;2),(\beta^2;4),(\beta^3;6) \},&\quad&
B_{3,4} &=& \{ (u_2;3),(\beta u_2;5) \},\\
B_{3,1} &=& \{ (u_1;2),(\beta u_1;4),(\beta^2 u_1;6) \},&\quad&
B_{3,5} &=& \{ (u_1 u_2;5) \}, \\
B_{3,2} &=& \{ ({u_1}^2;4),(\beta {u_1}^2;6) \},&\quad&
B_{3,6} &=& \{ ({u_2}^2;6) \}. \\
B_{3,3} &=& \{ ({u_1}^3;6) \},  
\end{array}
\]
and 
\begin{eqnarray*}
{\cal B}_{3} &=& 
\{ (1;0), (\beta;2), (\beta^2;4), (\beta^3;6), (u_1;2), (\beta u_1;4),
(\beta^2 u_1;6), ({u_1}^2;4),\\
& &\; (\beta {u_1}^2;6), ({u_1}^3;6), (u_2;3), (\beta u_2;5), 
(u_1 u_2;5), ({u_2}^2;6) \}.
\end{eqnarray*}
\item[(b)] Set ${\cal G} = {\cal B}_{3}$.
\item[(c)] Next, we apply derivatives to the first components of the 
pairs in ${\cal G}$.
\vskip 5pt
\noindent
Computation of $\ell$ for each pair of ${\cal G}$ gives
\[ 
\ell = 6,4,2,0,4,2,0,2,0,0,3,1,1,\; {\rm and} \; 0. 
\] 
Note that in this case all values for $\ell$ are integer.
Gathering the terms that come from applying the indicated number, 
$\ell,$ of partial derivatives with respect to $x$, gives 
\[
\!{\cal H}\!=\!\{0, \beta^3, u_{1,4x}, 
\beta u_{1,2x}, \beta^2 u_1, {u^2_{1,x}}, u_1 u_{1,2x}, 
\beta u_1^2, u_1^3, u_{2,3x}, \beta u_{2,x}, u_1 u_{2,x}, 
u_{1,x} u_2, u_2^2 \}. 
\]
\item[(d)] Removing from ${\cal H}$ the constant terms, and the terms 
that can be written as a $x-$derivative, or as a $x-$derivative up to 
terms retained earlier in the set ${\cal I}$, yields
\[ {\cal I} = \{ \beta^2 u_1, \beta {u_1}^2, {u_1}^3, {u_2}^2, u_{1,x} {u_2},
   {u_{1,x}}^2 \}. \]
\item[(e)] Combining these building blocks, the form of the density with 
rank 6 follows;
\[ \rho = c_1 \; \beta^2 u_1 + c_2 \; \beta {u_1}^2 + c_3 \; {u_1}^3 +
          c_4 \; {u_2}^2 +c_5 \; u_{1,x} {u_2} + c_6 \; {u_{1,x}}^2 . \]
\end{enumerate}
\noindent
Using Step 3 of the algorithm (illustrated in detail in the next example) 
we compute the density of rank 6 in the original variables:
$$ \rho = \beta u^2 - u^3 + v^2 + \alpha {u_x}^2. $$
\vskip 6pt
\noindent
Analogously, we computed densities of rank $R \le 6.$
The first four densities of (\ref{Boussys2}) are:
\[
\begin{array}{rclcrcl}
\rho_1 &=& u,& \quad\quad &
\rho_2 &=& v,\\ 
\rho_3 &=& u v,& \quad\quad &
\rho_4 &=& \beta u^2 - u^3 + v^2 + \alpha {u_x}^2.
\end{array} 
\]
After substitution of $\beta = 1$ into the densities above, one gets 
the densities of
(\ref{Boussys1}) even though initially this system was not uniform in rank.
This trick, which involves the introduction of one or more 
extra parameters with weights, can always be attempted if any equation 
in (\ref{givensystem}) lacks uniformity in rank.
}
\end{example}
\vskip 5pt
\noindent
\begin{example}
{\rm
Hirota and Satsuma (1981) proposed a coupled system of KdV equations, 
\begin{eqnarray}\label{HirSatpar}
u_t - 6 \alpha u u_x + 6 v v_x - \alpha u_{3x} &=& 0, \nonumber \\
v_t + 3 u v_x + v_{3x} &=& 0,
\end{eqnarray}
where $\alpha$ is a nonzero parameter. 
System (\ref{HirSatpar}) describes interactions of two long waves with 
different dispersion relations. It is known to be completely integrable for 
$\alpha=\frac{1}{2}.$

In our notation, (\ref{HirSatpar}) can be rewritten as
\begin{eqnarray}\label{prghirsatpar}
u_{1,t} - 6 \alpha {u_1} u_{1,x} + 6 u_2 u_{2,x} - 
\alpha u_{1,3x} &=& 0,
    \nonumber \\
u_{2,t} + 3 u_1 u_{2,x} + u_{2,3x} &=& 0, 
\end{eqnarray}
which has scaling properties, 
$ u_1 \sim u_2 \sim \frac{\partial^2{}}{\partial{x^2}}, \;
\frac{\partial{}}{\partial{t}} \sim \frac{\partial^3{}}{\partial{x^3}}, $ 
and a density of rank 4:
\[ \rho = c_1 \; {u_1}^2 + c_2 \; u_1 u_2 + c_3 \; {u_2}^2.\]
Thus far we used Steps 1 and 2 of the algorithm.
We now illustrate Step 3, which fixes the undetermined 
coefficients $c_i.$
\begin{enumerate}
\item[(a)] Compute $D_t \rho$ and replace all the mixed derivatives 
${(u_{i,t})}^{(j)}$  by using (\ref{prghirsatpar}). Then,
\begin{eqnarray*}
\!\!\!\!\!\!\!\!E \!\!\!\!&=&\!\!\!\! 
2 c_1 u_1 \!\left( 6 \alpha u_1 u_{1,x} - 
      6 u_2 u_{2,x} + \alpha u_{1,3x} \right) \! + 
c_2 u_2 \! \left( 6 \alpha u_1 u_{1,x} - 
      6 u_2 u_{2,x} + \alpha u_{1,3x} \right) \\ 
& & - c_2 u_1 \left( 3 u_1 u_{2,x} + u_{2,3x} \right)-
     2 c_3 u_2 \left( 3 u_1 u_{2,x} + u_{2,3x} \right) .
\end{eqnarray*}
\item[(b)] Apply the Euler
operator to get the linear system for the coefficients $c_1,c_2$ and $c_3$:
\[ {\cal S} = \{ (1+\alpha) c_2 = 0,\; 2 \; c_1+c_3 = 0 \}. \] 
\vskip 2pt
\noindent
\item[(c)] Obviously, ${\cal C} = \{ c_1,c_2,c_3 \}$ and $\cal S$
has one parameter, $\alpha.$
Thus, we search for compatibility conditions:
\begin{itemize}
\item Set $\boldmath c_1 = 1, $ which gives 
\[ \{ c_1 = 1, \; c_2 = 0, \; c_3 = -2 \} \]
as one of the solutions without any constraint on the parameter $\alpha$.
Since only $c_2 = 0,$ one has ${\cal Z} = \{c_2\}$ and 
$ {\cal C} = {\cal C} \cap {\cal Z} = \{ c_2 \},$ with $i' = 2$.
\item Set $\boldmath c_2 = 1.$ This leads to the compatibility 
condition $\alpha = -1$ and the solution
\[ \{ c_1 = - \frac{1}{2}\; c_3, \;\; c_2 = 1 \}. \]
Since ${\cal Z} = \{ \} $ and, consequently, 
${\cal C} = {\cal C} \cap {\cal Z} = \{ \},$ the procedure ends.
\end{itemize}
\end{enumerate}
Therefore, one gets {\it two densities} of rank 4, one without
any constraint on $\alpha,$ one with a constraint. 
In summary: 
\begin{itemize}
\item $\rho = {u_1}^2 - 2 \; {u_2}^2, \;$ and 
\vskip 4pt
\noindent
\item $\rho = - \frac{1}{2} c_3 {u_1}^2 + u_1 u_2 + c_3 {u_2}^2, \;$ with
compatibility condition $\alpha = -1.$
\end{itemize}
\vskip 2pt
\noindent
A search for densities for (\ref{HirSatpar}) of rank $R \le 8$ resulted in:
\vskip 6pt
\noindent
{\bf Rank 2:} There is no condition on $\alpha.$ 
One always has the density $ \rho = u. $
\vskip 10pt
\noindent
{\bf Rank 4:} At this level, two branches emerge:
\begin{enumerate}
\item Without condition on $\alpha,$ one obtains $ \rho = u^2 - 2 v^2.$
\item For $\alpha=-1$ one has
$\rho = u v - \frac{c}{2}\; (u^2 - 2 v^2),\;\; c {\rm \;\; is \;\; free}.$
\vskip 2pt
\noindent
Hence, for $\alpha=-1,$ there is a second independent conserved 
density: $\rho = u v.$
\end{enumerate}
{\bf Rank 6:} There is no condition on $\alpha.$ One obtains
\[ \rho= (1 + \alpha ) u^3 - 3 u v^2 - \frac{1}{2}(1+\alpha) {u_x}^2
+ 3 {v_x}^2. \]
{\bf Rank 8:} The system has conserved density
\[ \rho=u^4 - \frac{12}{5} u^2 v^2 + \frac{12}{5} v^4
- 2 u {u_x}^2 + \frac{1}{5} {u_{2x}}^2 + \frac{8}{5} v u_{x} v_{x}
-\frac{24}{5} u {v_x}^2 + \frac{8}{5} {v_{2x}}^2, \] 
provided that $\alpha = \frac{1}{2}.$\\
Therefore, $\alpha = \frac{1}{2}$ appears again as one tries to compute the 
density of rank 8. 
\newline\noindent 
This value of $\alpha$ leads to the only integrable 
case in the parameterized system (\ref{HirSatpar}).
\vskip 4pt
\noindent
For $\alpha=\frac{1}{2},$ we also computed the density of {\bf Rank 10}:
\begin{eqnarray*}
\rho &\!=\!& \!u^5 -\frac{20}{7} u^3 v^2 +\frac{20}{7} u v^4 
- 5 u^2 {u_x}^2 
+ \frac{10}{7} v^2 {u_x}^2 + u {u_{2x}}^2
- \frac{1}{14} {u_{3x}}^2 + \frac{40}{7} u v u_{x} v_{x} \\
&& - \frac{20}{7} u^2 {v_x}^2 
-\frac{80}{7} v^2 {v_x}^2 - \frac{24}{7} u_{2x} {v_{x}}^2 
- \frac{4}{7} v u_{2x} v_{2x} + \!\frac{40}{7} u {v_{2x}}^2 - 
 \frac{8}{7} {v_{3x}}^2.   
\end{eqnarray*}
}
\end{example}
\vskip 1pt
\noindent
\begin{example}
{\rm
Ito (1982) proposed the coupled nonlinear wave equation 
(Ablowitz and Clarkson, 1991):
\begin{eqnarray}\label{Itosys}
u_t + 6 u u_x + 2 v v_x + u_{3x} &=& 0, \nonumber\\
v_t + 2 v u_x + 2 u v_x &=& 0,
\end{eqnarray}
which differs from the Hirota-Satsuma system in the interaction and 
dispersion terms for $v.$ 
In the absence of $v,$ system (\ref{Itosys}) reduces to the KdV equation. 
It is a Hamiltonian system with infinitely many conservation laws. 
The scaling properties of the system (\ref{Itosys}) are 
\[ u \sim v \sim \frac{\partial^2{}}{\partial{x^2}},\quad
  \frac{\partial{}}{\partial{t}} \sim \frac{\partial^3{}}{\partial{x^3}}, 
\]
and the first five densities are:
\[
\begin{array}{rclcrcl}
\!\!\!\!\!\rho_1 &=& c_1 u + c_2 v, &\;\quad & \!\rho_2 &=& u^2 + v^2, \\
\\
\!\!\!\!\!\rho_3 &=& u^3 + u v^2 -\frac{1}{2} {u_x}^2, & \;\quad & 
\!\rho_4 &=& u^4 +\!\frac{6}{5} u^2 v^2 +\!\frac{1}{5} v^4 -\!2 u {u_x}^2 
+ \!\frac{1}{5} {u_{2x}}^2 - \!\frac{4}{5} v u_{x} v_{x}, 
\end{array}
\]
\begin{eqnarray*} 
\rho_5 &=& \!u^5 + \!\frac{10}{7} u^3 v^2 + \!\frac{3}{7} u v^4
- \!5 u^2 {u_x}^2 - \!\frac{5}{7} v^2 {u_x}^2 + \!u {u_{2x}}^2
- \!\frac{1}{14} {u_{3x}}^2 - \!\frac{20}{7} u v u_x v_x 
- \!\frac{2}{7} v^2 {v_x}^2 \\
& & + \frac{2}{7} u_{2x} {v_{x}}^2 + \frac{2}{7} v u_{2x} v_{2x}. 
\end{eqnarray*}
\noindent
To illustrate Step 3 of the algorithm in more detail,
we consider,
\begin{eqnarray}\label{Itosyspar}
u_t + 6 u u_x + 2 v v_x + u_{3x} &=& 0, \nonumber\\
v_t + \alpha ( u_x v + u v_x ) &=& 0,
\end{eqnarray}
which is a parameterized version of (\ref{Itosys}). 
The form of the density of rank 6 is
\[ \rho = c_1 \; {u}^3 + c_2 \; {u}^2 v + c_3 \; u {v}^2 +
          c_4 \; {v}^3 + c_5 \; {u_x}^2 + c_6 \; u_x v_x + c_7 \; {v_x}^2.
\] 
\noindent
We continue with Part (b) of Step 3. After applying the Euler operator,
the linear system $\cal S$ for the coefficients $c_1$ through $c_7$ is
\begin{eqnarray*}
{\cal S} &=& \{ (6-\alpha) \; c_2 = 0,\; c_1 - c_3 = 0, \; 
                2 \; c_2 - 3 \; \alpha \; c_4 = 0, \;
                c_1 + 2 \; c_5 = 0, \; c_3 + 2 \; c_5 = 0, \\ 
           & & c_6 = 0, \; c_3 + 2 \; c_5 - \alpha \; c_7 = 0, \;
                \alpha \; c_7 = 0, \; 2 \; c_2 + \alpha \; c_6 = 0, \;
                2 \; c_2 - 6 \; c_6 + \alpha \; c_6 = 0 \}. 
\end{eqnarray*} 
Obviously, ${\cal C} = \bigcup_{i=1}^{7} \{c_i\}$ and $\cal S$ has one 
parameter, $\alpha.$ Thus, we start the search for compatibility conditions. 
From the sixth and eighth equations in $\cal S$ we conclude 
that $ c_6 = c_7 = 0 $. Therefore, replace ${\cal C} $ by
$ {\cal C} \backslash \{ c_6, \; c_7 \} = \bigcup_{i=1}^{5} \{c_i\} $. 
\begin{itemize}
\item Set $\boldmath c_1 = 1 $ and solve $\cal S$, which gives 
\[ \{ c_1 = 1, \; c_2 = 0, \; c_3 = 1, \; c_4 = 0, \; c_5 = - \frac{1}{2}, \;
      c_6 = 0, \; c_7 = 0 \}, \]
without any constraint on the parameter $\alpha$.
Now, ${\cal Z} = \{c_2, \; c_4, \; c_6, \; c_7 \}$ and 
$ {\cal C} $ is replaced by 
$ {\cal C} \cap {\cal Z} = \{ c_2, \; c_4 \},$ with $i' = 2$.
\item Set $\boldmath c_2 = 1.$ This leads to an inconsistent 
system. Thus, ${\cal C} = {\cal C} \backslash \{ c_2 \} = \{ c_4 \},$ with 
$i' = 4$.
\item Set $\boldmath c_4 = 1.$ This implies that $\alpha = 0$. 
Hence, ${\cal C} = {\cal C} \backslash \{ c_4 \} = \{  \},$ 
and the procedure ends.
\end{itemize}
In summary: 
$
\displaystyle{ \rho = {u}^3 + u \; {v}^2 - \frac{1}{2} {u_x}^2 },
$ 
without any constraint on $\alpha$. 
This is the same density as for (\ref{Itosys}).

Computation of the density of rank 8 for (\ref{Itosyspar}) leads 
to the condition $\alpha = 2 $ and density $\rho_4$ listed above.
For $\alpha = 2,$ system (\ref{Itosyspar}) reduces to the
integrable case (\ref{Itosys}).
}
\end{example}
\section{Applications}

For systems with parameters, the algorithm described in Section 2 can be 
used to find the necessary conditions on the parameters so that the systems 
might have densities of fixed rank.
If for certain parameters a system has a large number of conserved 
densities, it is candidate to be completely integrable.
This is the major application of the algorithm, and also of our Mathematica
program {\bf condens.m}. 
The next examples illustrate the computer analysis of equations with 
parameters.

\subsection{Fifth-Order Korteweg-de Vries Equations}

Consider the family of fifth-order KdV equations,
\begin{equation} \label{KdV5par}
u_t + \alpha u^2 u_{x} + \beta u_x u_{2x} + \gamma u u_{3x} + u_{5x} = 0,
\end{equation}
where $\alpha,\beta,\gamma$ are nonzero parameters.
Special cases of (\ref{KdV5par}) are well known in the literature
(Fordy and Gibbons, 1980; Hirota and Ito, 1983; Kupershmidt and Wilson, 1981;
Satsuma and Kaup, 1977). 
Indeed, for $\alpha = 30, \beta = 20, \gamma = 10,$ equation (\ref{KdV5par})
reduces to the Lax (1968) equation.
The SK equation, due to Sawada and Kotera (1974), and also Dodd and 
Gibbon (1977), is obtained for $\alpha = 5, \beta = 5, \gamma = 5.$   
The KK equation, due to Kaup (1980) and Kupershmidt, corresponds to 
$\alpha = 20, \beta = 25, \gamma = 10,$ and the equation due to Ito (1980)
arises for $\alpha = 2, \beta = 6, \gamma = 3$.

The scaling properties of (\ref{KdV5par}) are such that 
\[ u \sim \frac{\partial^2}{\partial{x^2}},\quad
   \frac{\partial}{\partial{t}} \sim \frac{\partial^5}{\partial{x^5}}. \]
Using our algorithm, one easily computes the {\em compatibility conditions}
for the parameters $\alpha, \beta$ and $\gamma,$ so that (\ref{KdV5par}) 
admits a polynomial conserved density of fixed rank.
\vskip 3pt
\noindent
The results are:
\vskip 5pt
\noindent
{\bf Rank 2:} There are no conditions on the parameters. 
Indeed, equation (\ref{KdV5par}) can be written as a 
conservation law with density $\rho = u.$ 
\vskip 5pt
\noindent
{\bf Rank 4:} The equation has density $\rho = u^2$ provided that 
\begin{equation}\label{KdV5con4}
\beta = 2 \gamma.
\end{equation}
Only the Lax and Ito equations have this density.
\vskip 5pt
\noindent
{\bf Rank 6:} The condition 
\begin{equation}\label{KdV5con6}
\alpha = - \frac{1}{10} (2 {\beta}^2 - 7 \beta \gamma + 3 {\gamma}^2)
\end{equation}
guarantees the existence of the density of rank 6:
\[ \rho = (2 \beta-\gamma) u^3 - 15 {u_x}^2, \quad 
 \gamma \neq 2 \beta. \]
The Lax, SK and KK equations have this density.
\vskip 5pt
\noindent
{\bf Rank 8:} There are two branches:
\begin{enumerate}
\item If the condition (\ref{KdV5con4}) holds then
$$ 
\rho = \alpha u^4 - 6 \gamma u {u_x}^2 + 6 {u_{2x}}^2. 
$$
This branch covers the Lax and Ito cases.
\item If the condition
\begin{equation}\label{KdV5con82}
\alpha = - \frac{1}{45} (2 {\beta}^2 - 7 \beta \gamma - 4 {\gamma}^2 )
\end{equation}
holds, one has 
\[ 
\rho = {(2 \beta + \gamma )}^2 u^4 - 135 (2 \beta + \gamma) u {u_x}^2 + 
675  {u_{2x}}^2, \quad \gamma \neq -2 \beta. 
\]
This branch covers the SK, KK and Ito equations.
\end{enumerate}
{\bf Rank 10:} The conditions
\begin{equation}\label{KdV5con10}
\beta = 2 \gamma \;\;{\rm and}\;\; \alpha = \frac{3}{10} {\gamma}^2
\end{equation}
must hold in order to have the density
\[ 
\rho = \gamma^3 u^5 - 50 \gamma^2 u^2 {u_x}^2 + 100 \gamma u {u_{2x}}^2 
- \frac{500}{7} {u_{3x}}^2. 
\]
Only the Lax equation has this density.
Naturally, the following question arises:
What are the necessary conditions for the parameters 
$\alpha, \beta $ and $\gamma $ so that (\ref{KdV5par}) might have 
infinitely many polynomial conservation laws? 
\vskip 5pt
\noindent
{\bf Lax Case:} The Lax equation admits densities of rank 4 and 6. 
Combining the conditions (\ref{KdV5con4}) and (\ref{KdV5con6}) leads
to 
\begin{equation} \label{Lax5con}
\alpha = \frac{3}{10} \gamma^2 {\rm \;\;and\;\;} \beta = 2 \gamma,
\end{equation}
fixing $\alpha$ and $\beta$ in terms of $\gamma$ (the latter parameter can
always be scaled to 1).
The conditions in (\ref{Lax5con}) lead to the Lax equation, which is known to 
be completely integrable and has infinitely many conserved densities.
\vskip 5pt
\noindent
{\bf SK-KK Cases:} The SK and KK equations admit densities of rank 6 and 8.
Combining (\ref{KdV5con6}) and (\ref{KdV5con82}) gives
\begin{eqnarray} 
\alpha = \frac{1}{5} \gamma^2 &{\rm and}& \beta = \gamma,
\label{SK5con} \\
\alpha = \frac{1}{5} \gamma^2 &{\rm and}& \beta = \frac{5}{2} \gamma.
\label{KK5con}
\end{eqnarray}
The conditions in (\ref{SK5con}) lead to the SK equation, 
and those in (\ref{KK5con}) to the KK equation. Both equations are 
indeed integrable and have an infinite sequence of conserved densities. 
\vskip 5pt
\noindent
{\bf Ito Case:} The Ito equation admits only three densities.
Combining (\ref{KdV5con4}) and (\ref{KdV5con82}) gives
\begin{equation} \label{Ito5con}
\alpha = \frac{2}{9} \gamma^2 {\rm \;\;and\;\;} \beta = 2 \gamma.
\end{equation}
These conditions reduce (\ref{KdV5par}) to the Ito equation which is not 
integrable.
\vskip 2pt
\noindent
Using our code we were able to retrieve all known integrable cases 
in the family (\ref{KdV5par}), as given in e.g. Mikhailov {\it et al.\/} 
(1987). 
The Lax equation comes as no surprise since it is the fifth-order symmetry 
of the KdV equation (\ref{KdVeq}). 
Our program confirms that, apart from the KdV equation, there
are only two non-trivial integrable equations in the family 
(\ref{KdV5par}), namely the KK and SK equations. 
Some of the conserved densities are listed in Tables 1 and 2.
A proof for the gaps (with period three) between conservation laws for the
SK and KK equations is given in Sanders and Wang (1996).
Mathematical rigorous proofs for the existence of conservation laws of 
such evolution equations are given in Sanders and Wang (1995b, 1996).

\subsection{Seventh-Order Korteweg-de Vries Equations}

Consider the family of seventh-order KdV equations,
\begin{equation}\label{KdV7par}
u_t+a u^3 u_x+b {u_x}^3 +c u u_x u_{2x}+d u^2 u_{3x} +e u_{2x} u_{3x}+
f u_x u_{4x} + g u u_{5x} + u_{7x} = 0,
\end{equation}
where $a,b,c,d,e,f$ and $g$ are nonzero parameters. Special cases of 
(\ref{KdV7par}) are known in the literature. 
For $a = 252, b = 63, c = 378, d = 126, e = 63, f = 42, g = 21$, equation
(\ref{KdV7par}) reduces to the SK-Ito equation, due to Sawada and Kotera, 
and Ito (1980).
For $a = 140, b = 70, c = 280,d = 70, e = 70, f = 42, g = 14,$ equation
(\ref{KdV7par}) belongs to the KdV hierarchy studied by Lax (1968). 
For $a = 2016, b = 630, c = 2268,d = 504,$ 
$e = 252, f = 147, g = 42,$ 
the equation belongs to the Kaup-Kupershmidt hierarchy 
(Bilge, 1992; Gerdt, 1993).
The scaling properties of (\ref{KdV7par}) are
$ \displaystyle {
u \sim \frac{\partial^2{}}{\partial{x^2}},\quad
\frac{\partial{}}{\partial{t}} \sim \frac{\partial^7{}}{\partial{x^7}}. 
}
$
\vskip 1pt
\noindent
With {\bf condens.m}, we computed the {\em compatibility conditions}
for the parameters, so that conserved densities of fixed rank exist. 
The results are:
\vskip 3pt
\noindent
{\bf Rank 2:} $ \rho = u $ if $b = \frac{1}{2}(c-2 d)$.
True for both the Lax and SK-Ito equations.
\vskip 5pt
\noindent
{\bf Rank 4:} One has $ \rho = u^2 $ if
$b = c - 3 d {\rm \;\;and\;\;} e = 5 (f - 2 g);$ 
only true for Lax's equation.
\vskip 4pt
\noindent
{\bf Rank 6:} There are three branches. 
If one of the following three sets of conditions holds:
\begin{eqnarray*}
\!\!\!\!\!\!\!\!\!
a & = & \frac{1}{294} (28 b f -42 c f +120 f^3 -42 b g + 63 c g-720 f^2 g +
1350 f g^2-810 g^3), \\
d & = & \frac{5}{14} (2 f^2-9 f g +9 g^2) {\rm \;\;and\;\;}
e = \frac{14 c}{2 f-3 g}, \\
{\rm or } \\ 
a & = & \frac{1}{294} (28 b f -42 c f -40 f^3 -42 b g + 63 c g+240 f^2 g - 
450 f g^2+270 g^3), \\
d & = & - \frac{5}{42} (2 f^2-9 f g +9 g^2) {\rm \;\;and\;\;}
e = \frac{2}{3} \frac{(21 c+20 f^2-90 f g+90 g^2)}{2 f-3 g}, \\
{\rm or } \\
a & = & \frac{1}{42} (4 b f -6 c f+24 d f -6 b g +9 c g -36 d g), \\
e & = & \frac{14 c -14 d+10 f^2-45 f g+45 g^2}{2 f -3 g},
\end{eqnarray*}
then 
$$
\rho = (2 f - 3 g) u^3 - 21 {u_x}^2, \quad f \neq \frac{3}{2} g.
$$
{\bf Rank 8:}
\begin{eqnarray*}
& & \rho = 
(49 c+10 f^2-45 f g+20 g^2) u^4 - 252 (2 f-g)  u {u_x}^2 + 1764 {u_{2x}}^2, \\
& & c \neq -\frac{1}{49} (10 f^2 - 45 f g + 20 g^2),
\end{eqnarray*}
provided that
\begin{eqnarray*}
a & = & -\frac{1}{882}(49 c f+10 f^3-196 c g-85 f^2 g+200 f g^2-80 g^3 ),
\nonumber \\
b & = & \frac{1}{42}(14 c+2 f^2-9 f g+4 g^2), \quad 
d  = \frac{1}{42}(7 c-2 f^2+9 f g-4 g^2), {\rm \;\;and\;\;} \\
e & = & 2 f-g .
\end{eqnarray*}
\vskip 2pt
\noindent
Combining the conditions for {\bf Rank 2} through {\bf Rank 8} yields
\begin{eqnarray}
& & a= \frac{4}{147} g^3,\;\; b = \frac{1}{7} g^2,\;\; c = \frac{6}{7} g^2, 
\;\; d = \frac{2}{7} g^2,\;\; e = 3 g,\; f = 2 g, \label{SKIto7con} \\
{\rm or} \nonumber \\
& & a=\frac{5}{98} g^3,\;\; b=\frac{5}{14} g^2,\;\; c=\frac{10}{7} g^2,\;\;
d = \frac{5}{14} g^2,\;\; e= 5 g,\;\; f = 3 g ,\label{Lax7con} \\
{\rm or} \nonumber \\
& & a=\frac{4}{147} g^3,\;\; b=\frac{5}{14} g^2,\;\; c=\frac{9}{7} g^2,\;\;
d = \frac{2}{7} g^2,\;\; e= 6 g,\;\; f = \frac{7}{2} g. \label{Third7con} 
\end{eqnarray}
In each case all of the parameters are fixed in terms of $g,$
which could be scaled to 1.
The equations (\ref{SKIto7con}), (\ref{Lax7con}), and (\ref{Third7con})
belong to the SK-Ito, Lax and KK hierarchies, respectively, and they are 
all integrable (Bilge, 1992). 
Some of the conserved densities for the SK-Ito and Lax equations are listed 
in Table 3 for the choices $g=21$ and $g=14,$ respectively.

The conditions in (\ref{Third7con}) are satisfied for 
$a = 2016, b = 630,$
$c = 2268, d = 504,$ 
$e = 252, f = 147, g = 42,$ and the first 
five conserved densities of (\ref{KdV7par}), 
corresponding to the KK-hierarchy, then are: 
\begin{eqnarray*}
\rho_1 &=& u, 
\quad\quad\quad\quad 
\rho_2 = u^3 - \frac{1}{8} {u_x}^2, 
\quad\quad\quad\quad 
\rho_3 = u^4 -\frac{3}{4} u {u_x}^2 + \frac{1}{48} {u_{2x}}^2, \\
\\
\rho_4 &=& u^6 - \frac{35}{8} u^3 {u_x}^2 - \frac{31}{384} {u_x}^4 +
\frac{17}{32} u^2 {u_{2x}}^2 +\frac{37}{1152} {u_{2x}}^3 
- \frac{5}{192} u {u_{3x}}^2 + \frac{1}{2304} {u_{4x}}^2, \\
\\
\rho_5 &=& u^7 - \frac{63}{8} u^4 {u_x}^2 - \frac{287}{320} u {u_x}^4
+ \frac{161}{120} u^3 {u_{2x}}^2 + \frac{97}{640} {u_x}^2 {u_{2x}}^2 
+ \frac{737}{2880} u {u_{2x}}^3 \\
\\
& & - \frac{53}{480} u^2 {u_{3x}}^2 - \frac{133}{5760} u_{2x} {u_{3x}}^2
+ \frac{1}{240} u {u_{4x}}^2 - \frac{1}{17280} {u_{5x}}^2.
\end{eqnarray*}
Our results merely confirm the computer analysis of (\ref{KdV7par}) 
carried out with REDUCE by Bilge (1992) and Gerdt (1993).
\begin{table}[htbp]
\caption{Conserved Densities for Sawada-Kotera and Lax 5th-order Equations}
\label{table1}
\begin{center}
\begin{tabular}{||c|l|l||} 
\hline
& & \\
Density
& $\;\;\;\;\;\;$ Sawada-Kotera equation
& $\;\;\;\;\;\;\;\;\;\;\;\;\;\;\;$ Lax equation \\
& & \\ \hline
& & \\
$\rho_1$ & $u$ & $u$ \\
& & \\ \hline
& & \\
$ \rho_2$ & ---& $\frac{1}{2} u^2$ \\
& & \\ \hline
& & \\
$\rho_3 $ & $\frac{1}{3}  u^3- {u_x}^2 $
&  $\frac{1}{3} u^3 -\frac{1}{6} {u_x}^2 $ \\
& & \\ \hline
& & \\
$ \rho_4 $ & $ \frac{1}{4} u^4- \frac{9}{4}u {u_x}^2 +\frac{3}{4} {u_{2x}}^2$
& $ \frac{1}{4} u^4- \frac{1}{2}u {u_x}^2 +\frac{1}{20} {u_{2x}}^2$ \\
& & \\ \hline
& & \\
$\rho_5$ & ---  & $\frac{1}{5} u^5
-  u^2 {u_x}^2
+ \frac{1}{5} u {u_{2x}}^2
- \frac{1}{70} {u_{3x}}^2 $  \\
& & \\ \hline
& & \\
$\rho_6$ & $\frac{1}{6} u^6
- \frac{25}{4} u^3 {u_x}^2
- \frac{17}{8}  {u_x}^4
+ 6 u^2 {u_{2x}}^2$
  & $\frac{1}{6} u^6
- \frac{5}{3} u^3 {u_x}^2
- \frac{5}{36} {u_x}^4
+ \frac{1}2 u^2 {u_{2x}}^2 $  \\
& & \\
&$+ 2 {u_{2x}}^3 - \frac{21}{8} u {u_{3x}}^2
+ \frac{3}{8} {u_{4x}}^2 $
 & $+ \frac{5}{63} {u_{2x}}^3  - \frac{1}{14} u {u_{3x}}^2
+ \frac{1}{252} {u_{4x}}^2 $ \\
& & \\ \hline
& & \\
$\rho_7$ & $\frac{1}{7} u^7
- 9 u^4 {u_x}^2
- \frac{54}{5} u {u_x}^4
+ \frac{57}{5} u^3 {u_{2x}}^2$
  & $\frac{1}{7} u^7
- \frac{5}{2} u^4 {u_x}^2
- \frac{5}{6} u {u_x}^4
+ u^3 {u_{2x}}^2 $  \\
& & \\
&$+ \frac{648}{35} {u_x}^2 {u_{2x}}^2 + \frac{489}{35} u {u_{2x}}^3
- \frac{261}{35} u^2 {u_{3x}}^2 $
 & $+ \frac{1}{2} {u_x}^2 {u_{2x}}^2  + \frac{10}{21} u {u_{2x}}^3
- \frac{3}{14} u^2 {u_{3x}}^2 $ \\
& & \\
&$- \frac{288}{35} u_{2x} {u_{3x}}^2 + \frac{81}{35} u {u_{4x}}^2
- \frac{9}{35} {u_{5x}}^2$
 & $- \frac{5}{42} u_{2x} {u_{3x}}^2 + \frac{1}{42} u {u_{4x}}^2
- \frac{1}{924} {u_{5x}}^2$ \\
& & \\ \hline
& & \\
$\rho_8$ & ---  & $\frac{1}{8} u^8
- \frac{7}{2} u^5 {u_x}^2
- \frac{35}{12} u^2 {u_x}^4
 + \frac{7}{4} u^4 {u_{2x}}^2 $  \\
& & \\
& & $+ \frac{7}{2} u {u_{x}}^2 {u_{2x}}^2 + \frac{5}{3} u^2  {u_{2x}}^3
+ \frac{7}{24} {u_{2x}}^4 $ \\
& & \\
& & $ + \frac{1}{2} u^3 {u_{3x}}^2- \frac{1}{4} {u_{x}}^2 {u_{3x}}^2
- \frac{5}{6} u u_{2x} {u_{3x}}^2 $ \\
& & \\
& &$+ \frac{1}{12} u^2  {u_{4x}}^2 + \frac{7}{132} u_{2x} {u_{4x}}^2
- \frac{1}{132} u {u_{5x}}^2$ \\
& & \\
& & $ + \frac{1}{3432} {u_{6x}}^2 $ \\
& &  \\  \hline
\end{tabular}
\end{center}
\end{table}
\noindent
\begin{table}[htbp]
\caption{Conserved Densities for Kaup-Kupershmidt and Ito 5th-order Equations}
\label{table2}
\begin{center}
\begin{tabular}{||c|l|l||} 
\hline
& & \\
Density
& $\;\;\;\;\;\;$ Kaup-Kupershmidt equation
& $\;\;\;\;\;\;\;$ Ito equation \\
& & \\ \hline
& & \\
$\rho_1$ & $u$ & $u$ \\
& & \\ \hline
& & \\
$ \rho_2$ & ---& $\frac{1}{2} u^2$ \\
& & \\ \hline
& & \\
$\rho_3 $ & $\frac{1}{3} u^3 -\frac{1}{8} {u_x}^2 $
&  --- \\
& & \\ \hline
& & \\
$\rho_4 $ & $\frac{1}{4} u^4-\frac{9}{16}u {u_x}^2 +\frac{3}{64} {u_{2x}}^2$
& $ \frac{1}{4} u^4- \frac{9}{4}u {u_x}^2 +\frac{3}{4} {u_{2x}}^2$ \\
& & \\ \hline
& & \\
$\rho_5$ & ---  &--- \\
& & \\ \hline
& & \\
$\rho_6$ & $\frac{1}{6} u^6
- \frac{35}{16} u^3 {u_x}^2
- \frac{31}{256} {u_x}^4
+ \frac{51}{64}  u^2 {u_{2x}}^2$
  & ---  \\
& & \\
&$+\frac{37}{256} {u_{2x}}^3 - \frac{15}{128} u  {u_{3x}}^2
+ \frac{3}{512} {u_{4x}}^2 $
 &   \\
& & \\ \hline
& & \\
$\rho_7$ & $\frac{1}{7} u^7
- \frac{27}{8} u^4 {u_x}^2
- \frac{369}{320} u {u_x}^4
+ \frac{69}{40} u^3  {u_{2x}}^2$
  & ---  \\
& & \\
&$+ \frac{2619}{4480} {u_x}^2 {u_{2x}}^2 + \frac{2211}{2240} u  {u_{2x}}^3
- \frac{477}{1120} u^2 {u_{3x}}^2 $
 &  \\
& & \\
&$- \frac{171}{640} u_{2x}  {u_{3x}}^2 + \frac{27}{560} u {u_{4x}}^2
- \frac{9}{4480}  {u_{5x}}^2$
 & \\
& & \\ \hline
& & \\
$\rho_8$ & ---  & --- \\
& &  \\ \hline
$ \rho_9$ & $\frac{1}{9} u^9
- \frac{13}{2} u^6 {u_x}^2
- \frac{427}{32} u^3 {u_x}^4
- \frac{10431}{8960} {u_x^6} $
& ---  \\
& & \\
& $ + \frac{21}{4} u^5 {u_{2x}}^2 
+ \frac{12555}{448} u^2 {u_x}^2 {u_{2x}}^2 
+ \frac{2413}{224} u^3 {u_{2x}}^3 $
&  \\
& & \\
& $ + \frac{16461}{1792} {u_x}^2 {u_{2x}}^3
+ \frac{1641}{256} u {u_{2x}}^4 
- \frac{267}{112} u^4 {u_{3x}}^2 $
& \\
& & \\
& $ - \frac{3699}{896} u {u_x}^2 {u_{3x}}^2
- \frac{4383}{448}  u^2 u_{2x} {u_{3x}}^2
- \frac{76635}{19712} {u_{2x}}^2 {u_{3x}}^2 $
& \\
& & \\
& $ -\frac{18891}{19712} u_x {u_{3x}}^3
+ \frac{141}{224} u^3 {u_{4x}}^2 
+ \frac{8649}{39424} {u_x}^2 {u_{4x}}^2 $
& \\
& & \\
& $ + \frac{27639}{19712} u u_{2x} {u_{4x}}^2
+ \frac{2715}{39424}  {u_{4x}}^3 
- \frac{927}{9856} u^2 {u_{5x}}^2 $ 
& \\
& & \\
& $ - \frac{2943}{39424} u_{2x} {u_{5x}}^2
+ \frac{9}{1232}  u {u_{6x}}^2 
- \frac{9}{39424} {u_{7x}}^2 $
& \\
& & \\ \hline
\end{tabular}
\end{center}
\end{table}
\noindent
\begin{table}[htbp]
\caption{Conserved Densities for Sawada-Kotera-Ito and 
Lax 7th-order Equations}
\label{table3}
\begin{center}
\begin{tabular}{||c|l|l||}
\hline
& & \\
Density
& $\;\;\;\;\;\;$ Sawada-Kotera-Ito equation
& $\;\;\;\;\;\;\;\;\;\;\;\;\;\;\;$ Lax equation \\
& & \\ \hline
& & \\
$\rho_1$ & $u$    & $u$ \\
& & \\ \hline
& & \\
$ \rho_2$ & ---   & $ \frac{1}{2} u^2$ \\
& & \\ \hline
& & \\
$\rho_3 $ & $\frac{1}{3} u^3 - \frac{1}{3} {u_x}^2 $
       &  $ \frac{1}{3} u^3 - \frac{1}{6} {u_x}^2 $ \\
& & \\ \hline
& & \\
$ \rho_4 $ & $\frac{1}{4} u^4 -\frac{3}{4} u {u_x}^2 +\frac{1}{12} {u_{2x}}^2$
       & $ \frac{1}{4} u^4 - \frac{1}{2} u {u_x}^2 + \frac{1}{20}{u_{2x}}^2$ \\
& & \\ \hline
& & \\
$\rho_5$ & --- & $\frac{1}{5} u^5 - u^2 {u_x}^2 + \frac{1}{5} u {u_{2x}}^2
- \frac{1}{70} {u_{3x}}^2 $  \\
& & \\ \hline
& & \\
$\rho_6$ & $ \frac{1}{6} u^6 - \frac{25}{12} u^3 {u_x}^2
- \frac{17}{72} {u_x}^4 + \frac{2}{3} u^2 {u_{2x}}^2 $
     & $ \frac{1}{6} u^6 - \frac{5}{3} u^3 {u_x}^2 - \frac{5}{36} {u_x}^4
     + \frac{1}{2} u^2 {u_{2x}}^2 $ \\
& & \\
& $
+\frac{2}{27} {u_{2x}}^3 - \frac{7}{72} u {u_{3x}}^2+\frac{1}{216} {u_{4x}}^2$
    & $+\frac{5}{63} {u_{2x}}^3 -\frac{1}{14} u {u_{3x}}^2 +
      \frac{1}{252} {u_{4x}}^2 $  \\
& & \\ \hline
& & \\
$\rho_7$ & $\frac{1}{7} u^7 - 3 u^4 {u_x}^2 - \frac{6}{5} u {u_x}^4 
+ \frac{19}{15} u^3  {u_{2x}}^2 $
& $ \frac{1}{7} u^7 - \frac{5}{2} u^4 {u_x}^2 - \frac{5}{6} u {u_x}^4
+ u^3  {u_{2x}}^2 $ \\
& & \\
& $ + \frac{24}{35} {u_x}^2 {u_{2x}}^2 + \frac{163}{315} u {u_{2x}}^3
- \frac{29}{105} u^2 {u_{3x}}^2$
& $ +\frac{1}{2} {u_x}^2 {u_{2x}}^2 + \frac{10}{21} u {u_{2x}}^3
- \frac{3}{14} u^2 {u_{3x}}^2 $ \\
& & \\
& $-\frac{32}{315} u_{2x} {u_{3x}}^2 + \frac{1}{35} u {u_{4x}}^2 
- \frac{1}{945} {u_{5x}}^2$
 & $ -\frac{5}{42} u_{2x} {u_{3x}}^2 + \frac{1}{42} u {u_{4x}}^2
- \frac{1}{924} {u_{5x}}^2 $ \\
& & \\ \hline
& & \\
$\rho_8$ & --- & $ \frac{1}{8} u^8 - \frac{7}{2} u^5 {u_x}^2
-\frac{35}{12} u^2 {u_x}^4 + \frac{7}{4}  u^4 {u_{2x}}^2 $ \\
& & \\
& & $ + \frac{7}{2} u {u_x}^2 {u_{2x}}^2 + \frac{5}{3} u^2 {u_{2x}}^3 +
\frac{7}{24} {u_{2x}}^4 $ \\
& & \\
& & $ - \frac{1}{2} u^3 {u_{3x}}^2 -\frac{1}{4} {u_x}^2 {u_{3x}}^2
-\frac{5}{6} u u_{2x} {u_{3x}}^2 $ \\ 
& & \\
& & $ + \frac{1}{12} u^2 {u_{4x}}^2 + \frac{7}{132} u_{2x} {u_{4x}}^2
- \frac{1}{132} u {u_{5x}}^2 $ \\
& & \\
& & $ + \frac{1}{3432} {u_{6x}}^2$ \\
& & \\ \hline
\end{tabular}
\end{center}
\end{table}
\section{More Examples}

\subsection{Nonlinear Schr\"{o}dinger Equation}

The nonlinear Schr\"{o}dinger (NLS) equation (Ablowitz and Clarkson, 1991),
\begin{equation}\label{NLSeq}
i q_t - q_{2x} + 2 {|q|}^2 q =0,
\end{equation}
arises as an asymptotic limit of a slowly varying dispersive wave envelope 
in a nonlinear medium, and as such has significant applications in 
nonlinear optics, water waves, and plasma physics.
Equation (\ref{NLSeq}) is known to be completely integrable,
and together with the ubiquitous KdV equation (\ref{KdVeq}), 
is one of the most studied 
soliton equations. 
\newline
\noindent
There are two different ways to compute conserved densities for (\ref{NLSeq}).
\vskip 3pt
\noindent
{\bf Method 1}$\;\;$ One can replace (\ref{NLSeq}) by 
\begin{eqnarray}\label{NLSsys}
u_t - v_{2x}+2 v (u^2+v^2) &=& 0, \nonumber \\
v_t + u_{2x}-2 u (u^2+v^2) &=& 0,
\end{eqnarray}
where $q = u + i v.$ The scaling properties are such that
\[u \sim v \sim \frac{\partial{}}{\partial{x}},\quad
  \frac{\partial{}}{\partial{t}} \sim \frac{\partial^2{}}{\partial{x^2}}. \]
We computed seven conserved densities with our program, of which the first 
six are:
\[
\begin{array}{rclcrcl}
\!\!\!\!\!\!\rho_1 &=& u^2+v^2, 
&\;\;\quad\quad & 
\rho_2 &=& v u_x, \\
\!\!\!\!\!\!\rho_3 &=& u^4+2 u^2 v^2+v^4+{u_x}^2+{v_x}^2, 
&\;\;\quad\quad &
\rho_4 &=& u^2 v u_x + \frac{1}{3} v^3 u_x + \frac{1}{6} v_x u_{2x}, 
\end{array}
\]
\begin{eqnarray*}
\rho_5 &=& u^6 + 3 u^4 v^2 + 3 u^2 v^4 + v^6 + 5 u^2 {u_x}^2 + 3 v^2 {u_x}^2 
+ 3 u^2 {v_x}^2 + 5 v^2 {v_x}^2 + 4 \;u v u_x v_x \\
& &  + \frac{1}{2} {u_{2x}}^2 + \frac{1}{2} {v_{2x}}^2,\\
\\
\rho_6 &=& u^4 v u_x + \frac{2}{3} u^2 v^3 u_x + \frac{1}{5} v^5 u_x
+ \frac{1}{3} u {u_x}^2 v_x + \frac{1}{3} u^2 u_{2x} v_x 
+ \frac{1}{3} v^2 u_{2x} {v_x} + \;\frac{1}{3} v u_x {v_x}^2 \\
& &  + \frac{1}{30} u_{3x} v_{2x}. 
\end{eqnarray*}
\vskip 1pt
\noindent
{\bf Method 2}$\;\;$ One could consider $q$ and $q^{*}$ as independent 
variables and add the complex conjugate of (\ref{NLSeq}) to the NLS equation. 
Absorbing $i$ in the scale of $t$ then yields: 
\begin{eqnarray}\label{NLSsys2}
q_t - q_{2x} + 2 q^2 q^{*} &=& 0, \nonumber \\
q_t^{*} + q_{2x}^{*} - 2 q^{*2} q &=& 0.
\end{eqnarray}
According to the procedure described at the end of Step 1 in Section 2.3, 
our program computes the constraints
\[
w(q) = 2 - w(q^{*}), \quad\quad w(q^{*}) = w(q^{*}), 
\quad w(\frac{\partial{}}{\partial{t}}) = 2,
\]
and sets the left hand sides of the first two equal to one.
Then, it solves the equations
\[
1 = 2 - w(q^{*}), \quad\quad 1 = w(q^{*}),
\]
piece by piece. Both lead to the solution $w(q^{*}) = 1.$ 
Hence, the program continues with $w(q) = w(q^{*}) = 1.$ 
\newline
\noindent
For rank 2 through rank 6 
our program produces the conserved densities:
\[
\begin{array}{rclcrcl}
\!\!\!\rho_1 &=& q q^{*}, & \quad\quad & \rho_2 &=& q^{*} q_x,\\
\\
\!\!\!\rho_3 &\!=\!& q^2 q^{*2} + q_x q_x^{*}, & \quad\quad &
\rho_4 &\!=\!& q q^{*2} q_x + \frac{1}{3} q_{2x} q_x^{*}, 
\end{array}
\]
\[
\!\!\!\!\!\!\!\;\;\rho_5 \;\;=\;\; q^3 q^{*3} + \frac{1}{2} q^{*2} q_x^2 
+ 4 q q^{*} q_x q_x^{*} + \frac{1}{2} q^2 q_x^{*2} 
+ \frac{1}{2} q_{2x} q_{2x}^{*}.
\]
Obviously, these two sets of conservation laws are connected 
(but not piece by piece) via a simple change of variables: 
$ u = \frac{1}{2} (q + q^{*}), v = \frac{1}{2i} (q - q^{*}). $

The second method has the advantage that the conserved densities are
expressed in the original variable $q$ and it conjugate $q^{*}.$ 
On the other hand, the conserved densities from Method 2 may not be
independent. 

\subsection{Non-dispersive Long Wave System}

The non-dispersive long wave equations (Kupershmidt, 1985a) 
\begin{eqnarray}\label{Longwavesys}
u_t +  v u_x + u v_x &=& 0, \nonumber \\
v_t + u_x  + v v_x  &=& 0,
\end{eqnarray}
is another example of an integrable system.

We use this example to illustrate how our code determines a free weight.
Indeed, for (\ref{Longwavesys}),  
$ w(u) = 2 w(v), \; w(v)= w(v), \;
w(\frac{\partial{}}{\partial{t}}) = 1 + w(v), $ 
with $w(v)$ as the only free weight. 
As described in Step 1 of the algorithm, the program sorts the right hand
sides of these constraints, sets their left hand sides equal to one, 
and proceeds with solving
\[
1 = w(v), \quad\quad
1 = 2 w(v), \quad\quad
1 = w(v) + 1,
\]
one by one. That leads to the choices $w(v) = 1, w(v) = \frac{1}{2}, $ and 
$w(v) = 0.$
Since the latter is zero it is incremented by one to get $w(v) = 1.$
\vskip 2pt
\noindent
The first choice, $w(v) =1,$ gives 
$w(u) = w(\frac{\partial{}}{\partial{t}}) = 2 .$ 
The second choice, $w(v) =\frac{1}{2},$ gives 
$w(u) = 1, w(\frac{\partial{}}{\partial{t}}) = \frac{3}{2}.$
Another valid choice (not considered by our program) would be 
$w(v) =\frac{1}{4},$ $w(u) = \frac{1}{2},$ and 
$w(\frac{\partial{}}{\partial{t}}) = \frac{5}{4}.$ Obviously, there
are infinitely many fractional choices for $w(v). $
Recall that fractional weights and ranks are indeed allowed.  

The program continues automatically with the smallest integer choice, 
$w(v) =1.$ Hence, $ w(u) = 2, $ or, symbolically,  
\[ 
u \sim {\partial^2 \over \partial x^2} \;\;\; {\rm and} \;\;\;
v \sim {\partial \over \partial x}.
\]
For rank one through eight we obtained the following densities: 
\[
\begin{array}{rclcrcl}
\rho_1 & = &  v, & \quad\quad &
\rho_2 & = &  u, \\ 
\rho_3 & = &  u v, & \quad\quad &
\rho_4 & = &  u^2 + u v^2,\\
\rho_5 & = &  u^2 v + \frac{1}{3} u v^3 ,& \quad\quad &
\rho_6 & = &  u^3 + 3 u^2 v^2 + \frac{1}{2} u v^4, \\
\rho_7 & = &  u^3 v + u^2 v^3 + \frac{1}{10} u v^5,& \quad\quad & 
\rho_8 & = &  u^4 + 6 u^3 v^2 + 3 u^2 v^4 + \frac{1}{5} u v^6.
\end{array}
\]
This set of densities remains the same for any valid choice of
$w(v)$. Indeed, we could have computed the conserved density
$ \rho = u v $ with different choices of the free weight $w(v).$
This $\rho$ has rank $R = w(u) + w(v) = 3 w(v).$ 
If we choose $w(v) = \frac{1}{2}$ then $ \rho = u v$ has rank 
$R = \frac{3}{2}.$ 
On the other hand, for the choice $w(v) = \frac{1}{4},$ this density
has rank $R = \frac{3}{4}.$ 
In conclusion, choosing $w(v)$ differently does not affect the form of the 
desired densities, provided the rank of $\rho$ is also adjusted appropriately.
Due to the superposition principle, the same argument can be made if 
$\rho$ is the sum of many terms, involving $x-$derivatives of the
dependent variables. 


\subsection{Three-Component Korteweg-de Vries Equation}

Consider the 3-component extension of the KdV equation (Kupershmidt, 1985b),
\begin{eqnarray}\label{3cKdV}
u_t -6 u u_x +2 v v_x+2 w w_x-u_{3x} &=& 0, \nonumber \\
v_t-2 v u_x-2 u v_x &=& 0, \\
w_t -2 w u_x-2 u w_x &=& 0, \nonumber
\end{eqnarray}
which can be written as a bi-Hamiltonian system with 
infinitely many conservation laws.
The scaling properties indicate that 
\[
 u \sim v \sim w \sim {\partial^2 \over \partial x^2}, \quad
 \frac{\partial{}}{\partial{t}} \sim \frac{\partial^3{}}{\partial{x^3}},
\]
and the first four densities for (\ref{3cKdV}) are:
\begin{eqnarray*}
\rho_1 &=& c_1 u +c_2 v +c_3 w, \quad 
\rho_2 = u^2-v^2-w^2, \quad
\rho_3 = u^3 - u v^2 - u w^2 - \frac{1}{2}{u_x}^2, \\
\\
\rho_4 &=& u^4 - \frac{6}{5} u^2 v^2 +\!\frac{1}{5} v^4 
- \!\frac{6}{5} u^2 w^2 +\!\frac{2}{5} v^2 w^2 + \! \frac{1}{5} w^4 
- \! 2 u {u_x}^2 \! + \!\frac{1}{5} u_{2x}^2 \!
+ \!\frac{4}{5} v u_x v_x \!+\!\frac{4}{5} w u_x w_x.
\end{eqnarray*}
Obviously, from $\rho_1$ we can see that $u, v$ and $w$ are independent
conserved densities.
\section{Using the Program condens.m}

We now describe the features, scope and limitations of our program 
{\bf condens.m}, which is written in {\it Mathematica} syntax (Wolfram, 1996).
The program {\bf condens.m} has its own {\it menu\/} interface with 
30 samples of data files. 
Users are assumed to have access to {\it Mathematica} (version 2.0 or higher). 
The code {\bf condens.m} and the data files should be put in one directory. 

\subsection{The Menu Interface}

\noindent
After {\it Mathematica\/} comes up with `In[1]:=', type 
\begin{verbatim} 
In[1]:= <<condens.m 
\end{verbatim}
to read in the code {\bf condens.m} and start the program. 
Via its menu interface, the program will automatically 
prompt you for answers.
\begin{example}
{\rm Let us compute the density of rank 4 (provided it exists) 
for the Drinfel'd-Sokolov system (Ablowitz and Clarkson, 1991):
\begin{eqnarray} \label{drinsoko}
& & u_t + 3 v v_x = 0, \nonumber \\
& & v_t + 2 v_{3x} + 2 u v_x + u_x v = 0. 
\end{eqnarray}
Since this example is in the menu, start the program and pick entry 25 
from the menu.}  
\begin{verbatim}
  *** MENU INTERFACE ***  (page: 3)
-------------------------------------------
 21) MVDNLS System (d_mvdnls.m)
 22) Kaup System-parameterized (d_pkaup.m)
 23) Kaup System (d_kaup.m)
 24) Kaup-Broer System (d_broer.m)
 25) Drinfel'd-Sokolov System (d_soko.m)
 26) Dispersive Long Wave System (d_disper.m)
 27) Non-dispersive Long Wave System (d_nodisp.m)
 28) 3-Component KdV System (d_3ckdv.m)
 29) 2-Component Nonlinear Schrodinger Equation (d_2cnls.m)
 30) Boussinesq System (d_bous.m)
 nn) Next Page
 tt) Your System
 qq) Exit the Program
------------------------------------------
ENTER YOUR CHOICE: 25
Enter the rank of rho: 4
Enter the name of the output file: drisokr4.o

*********************************************************

          WELCOME TO THE MATHEMATICA PROGRAM               
           by UNAL GOKTAS and WILLY HEREMAN                
       FOR THE COMPUTATION OF CONSERVED DENSITIES.
        Version 3.0 released on February 24, 1997          
                    Copyright 1996                            

*****************************************************************
.
.
*****************************************************************

                                       2
The normalized density rho[x,t] is : u 
                                      2

******************************************************************

                                          2         2
The corresponding flux j[x,t] is:  2 u  u   - 2 u     + 4 u  u 
                                      1  2       2,x       2  2,xx

******************************************************************

Result of explicit verification (rho_t + J_x) = 0

******************************************************************
\end{verbatim}
\noindent
{\rm
At the end of computation, the normalized density and flux are available.
In the absence of parameters, both are normalized according to the 
coefficient of the first term in the density. 
If there are parameters, the common denominators that have been multiplied 
through are shown. To see the density and flux in standard 
Mathematica notation, type \verb|rho[x,t]| or \verb|j[x,t]|.}
{\rm However, type \verb|pdeform[rho[x,t]]| or \verb|pdeform[j[x,t]]|
to see them in subscript notation.}

{\rm Note that the form of the densities $ \rho $ is not unique.
Densities can always be integrated by parts to obtain {\it equivalent}
forms, modulo total derivatives. 
In {\it Mathematica} version 2.2, equivalent forms can be obtained via the 
command \verb|Integrate[rho[x,t],x]|.}
\end{example}
\subsection{Preparing Data Files}

To test systems that are not in the menu, prepare a data file in
the format of our data files.
Of course, the name for a new data file should not coincide with any name 
already listed in the menu, unless you intended to modify those data files. 

\begin{example}
{\rm
For the parameterized Hirota-Satsuma system (\ref{HirSatpar}) the data 
file reads:
}
\begin{verbatim}
(* start of data file d_phrsat.m *)

debug = False;

(* Hirota-Satsuma System with one parameter *)

eq[1][x,t]=D[u[1][x,t],t]-aa*D[u[1][x,t],{x,3}]-
           6*aa*u[1][x,t]*D[u[1][x,t],x]+6*u[2][x,t]*D[u[2][x,t],x];

eq[2][x,t]=D[u[2][x,t],t]+D[u[2][x,t],{x,3}]+3*u[1][x,t]*D[u[2][x,t],x];

noeqs = 2;
name = "Hirota-Satsuma System (parameterized)";
parameters = {aa};
weightpars = {};

(* user can supply the rank of rho and a name for the output file *)
(* rhorank = 6; *)
(* myfile = "phrsatr6.o; *)

(* user can give the weights of u[1], u[2] and partial t *)
(* weightu[1]=2; weightu[2]=2; weight[t]=3; *)

(* user can give the form of rho. Here, for density of rank 6: *)
(* formrho[x,t]={c[1]*u[1][x,t]^3+c[2]*u[1][x,t]*u[2][x,t]^2+
                 c[3]*D[u[1][x,t],x]^2+c[4]*D[u[2][x,t],x]^2}; *)

formrho[x,t] = {};

(* end of data file d_phrsat.m *)
\end{verbatim}
\end{example}
\vskip 2pt 
\noindent
{\rm
Explanation of the lines in the data file: 
\vskip 4pt 
\noindent
\verb|debug = False;|
\vskip 3pt 
\noindent
No trace of internal computations, otherwise, set it {\it True}.
\vskip 6pt
\noindent
\verb|eq[k][x,t] = ...;|
\vskip 3pt 
\noindent
Give the $k^{th}$ equation of the system in {\it Mathematica} notation. 
Note that there is no \verb|== 0| at the end.
\vskip 6pt
\noindent
\verb|noeqs = 2;|
\vskip 3pt 
\noindent
Specify the number of equations in the system.
\vskip 6pt
\noindent
\verb|name = "Hirota-Satsuma System (parameterized)";|
\vskip 3pt 
\noindent
Pick a short name for the system. The quotes are essential.
\vskip 6pt
\noindent
\verb|parameters = {aa};|
\vskip 3pt 
\noindent
Give the list of the parameters in the system.
If there are none, set {\it parameters = \{ \}}.
\vskip 5pt
\noindent
\verb|weightpars = {};|
\vskip 3pt 
\noindent
Give the list of the parameters that are assumed to have weights. 
Note that weighted parameters are {\it not} listed in {\em parameters}, 
which is the list of parameters without weight.
\vskip 6pt
\noindent
\verb|(* rhorank = 6; *)|
\vskip 3pt 
\noindent
Optional; give the desired rank of the density. Useful if you want 
to work with the program less interactively (in batch mode).
\vskip 6pt
\noindent
\verb|(* myfile = "hirsatr6.o; *)|
\vskip 3pt 
\noindent
Optional; give a name of the output file. Useful for less 
interactive use of the program.
\vskip 6pt
\noindent
\verb|(* weightu[1]=2; weightu[2]=2; weight[t]=3; *)|
\vskip 3pt
\noindent
Optional; specify a choice for {\it some or all} of the weights.
The program then skips the computation of the weights, but
still checks for consistency. Particularly useful if there are several 
free weights and non-interactive use is preferred.
\vskip 1pt
\noindent
\verb|formrho[x,t] = {};|
\vskip 1pt 
\noindent
The program will {\it compute} the form of $ \rho $ of the given rank.
\vskip 1pt
\noindent
\begin{verbatim}
formrho[x,t]={c[1]*u[1][x,t]^3+c[2]*u[1][x,t]*u[2][x,t]^2+
             c[3]*D[u[1][x,t],x]^2+c[4]*D[u[2][x,t],x]^2};
\end{verbatim}
\vskip 1pt
\noindent
Alternatively, one could give a form of $ \rho $ (here for rank 6).
The density must be given in expanded form and with coefficients c[i]. 
The braces are essential. 
If $ \rho $ is given, the program skips both the computation 
of the weights and the form of the density. 
Instead, the code uses what is given and computes the coefficients c[i].  
This option allows one, for example, to test densities from the literature.
\vskip 6pt
\noindent
Anything within \verb|(*| and \verb|*)| are comments, and ignored by
{\it Mathematica}.
\vskip 6pt
\noindent
Once the data file is ready, run it via the choice ``\verb|tt) Your System|"
in the menu.
}
\subsection{Scope and Limitations}

Our program can handle PDEs that can be written as systems of 
evolution equations. The evolution equations must be polynomials in the 
dependent variables (no integral terms). Only two independent variables
($x$ and $t$) are allowed.
No terms in the evolution equations should {\it explicitly} depend on
$x$ and $t.$

Theoretically, there is no limit on the number of evolution equations.
In practice, for large systems, the computations may take a long time or 
require a lot of memory. 
The computational speed depends primarily on the amount of memory.

The program only computes polynomial conserved densities in the 
dependent variables and their derivatives, without explicit dependencies 
on $x$ and/or $t.$

By design, the program constructs only densities that are uniform in rank. 
The uniform rank assumption for the monomials in $\rho$ allows one to 
compute {\it independent} conserved densities piece by piece, 
without having to split linear combinations of conserved densities. 
Due to the superposition principle, a linear combination of conserved
densities of unequal rank is still a conserved density. 
This situation arises frequently when parameters with weight are 
introduced in the PDEs.

The input systems may have one or more parameters, which are assumed to 
be nonzero. If a system has parameters, the program will attempt to 
compute the compatibility conditions for these parameters such that 
densities (of a given rank) exist.

The assumption that all parameters in the given evolution equation must 
be nonzero is necessary.
As a result of setting parameters to zero in a given system of evolution 
equations, the weights and therefore the rank of $\rho$ might change. 

In general, the compatibility conditions for the parameters could be 
highly nonlinear, and there is no general algorithm to solve them. 
The program automatically generates the compatibility conditions, 
and solves them for parameters that occur linearly (see Section 3.2). 
Gr\"obner basis techniques could be used to reduce complicated 
nonlinear systems into equivalent, yet simpler, non-linear systems. 
For PDEs with parameters and when the system for the coefficients $c_i$
is complicated, the program saves that system and its coefficient matrix, 
etc., in the file {\it worklog.m}. 
Independent from the program, the worklog files 
can later be analyzed with Mathematica functions. 

The assumption that the evolution equations are uniform in rank is not 
very restrictive. If the uniform rank condition is violated, the user 
can introduce one or more parameters with weights.
This also allows for some flexibility in the form of the densities.  
Although built up with terms that are uniform in rank, the densities do not 
have to be uniform in rank with respect to the dependent variables 
{\it alone}. This is illustrated in Example 2.2.

Conversely, when the uniform rank condition {\it is} fulfilled, the 
introduction of extra parameters (with weights) in the given PDE 
leads to a linear combination of conservation laws, not to new ones.

In cases where it is not clear whether or not parameters with weight
should be introduced, one should start with parameters without weight.
If this causes incompatibilities in the assignment of weights (due to 
non-uniformity), the program may provide a suggestion. Quite often, 
it recommends that one or more parameters be moved from the list of
{\it parameters} into the list {\it weightpars} of weighted parameters.

For systems with two or more free weights, the user will 
be prompted to enter values for the free weights. If only one
weight is free, the program will automatically compute some choices for 
the free weight, and eventually continue with the lowest integer or 
fractional value (see Examples 4.1 and 4.2). 
The program selects this value for the free weight; it is just one choice 
out of possibly infinitely many.
If the algorithm fails to determine a suitable value, the user will be
prompted to enter a value for the free weight. 

Negative weights are not allowed, except for $w(\frac{\partial}{\partial t}).$
If  $w(\frac{\partial}{\partial t}) < 0, $ the program will give a warning, 
but continue with the computations.
Zero weights are allowed, but at least one of the dependent variables
must have positive weight. The code checks these conditions, and if they 
are violated the computations are aborted. 
Note that {\it fractional weights} and densities of {\it fractional rank} 
are permitted.

Our program is a tool in the search of the first half-dozen 
conservation laws. An existence proof (showing that there are
indeed infinitely many conservation laws) must be done analytically, 
e.g. by explicitly constructing the recursion operator 
(Kodama, 1985; Sanders and Wang, 1995b, 1996) that connects conserved 
densities, or by computing high-order symmetries with Lie symmetry 
software (Hereman, 1996). 
If our program succeeds finding a large set of independent conservation laws, 
there is a good chance that the system of PDEs has infinitely many 
conserved densities and that the recursion operator could be 
constructed explicitly. 
If the number of conservation laws is 3 or less, most likely the PDEs are
not integrable, at least not in that coordinate representation.

\section{Other Software Packages}
This section gives a review of other symbolic software for the computation 
of conserved densities, together with recent developments in this area.
\subsection{SYMCD}

The program {\bf SYMCD}, written by Ito (1994), is an improved version of 
{\bf CONSD} by Ito and Kako (1985). Both programs are in REDUCE. 

Similar to our program, {\bf SYMCD} uses scaling properties to 
compute polynomial conserved densities for systems of any number 
of evolution equations with uniform rank. 
{\bf CONSD} had a limit on the number of evolution equations. 
This limitation was removed in {\bf SYMCD}. 
Evolution equations must be polynomial in the dependent variables and 
their derivatives, and variables with negative weight are not allowed. 
With REDUCE 3.5 on an IBM Risc 6000 work station, we tested the version 
of {\bf SYMCD} released in 1996.
Our test cases included equations (\ref{KdVeq}), (\ref{Boussys2}), 
(\ref{HirSatpar}), (\ref{KdV5par}) and (\ref{3cKdV}).

For systems with or without parameters, {\bf SYMCD} gives the same conserved 
densities as our program (up to terms that can be converted via 
integration by parts).

However, {\bf SYMCD} does not properly handle systems with parameters.
It stops after generating the necessary conditions on the parameters, 
which must be analyzed separately. 
Such analyses revealed that the conditions do not always lead to a density of 
fixed rank. Indeed, in solving for the undetermined coefficients, 
{\bf SYMCD} considers all possible branches in the solution, 
irrespective of whether or not these branches lead to a conserved density,
as confirmed by Ito (1996).
Another major difference is that parameters with weights are not allowed
in {\bf SYMCD}, which restricts the scope of {\bf SYMCD} to systems with 
uniform rank.
In conclusion, our code is more sophisticated than {\bf SYMCD}
in handling systems with parameters and systems of non-uniform rank.
For more information contact Ito at
ito@puramis.amath.hiroshima-u.ac.jp. 

\subsection{DELiA}

The PC package {\bf DELiA} for {\it Differential Equations with Lie Approach}, 
developed in the period 1989-1991, is an outgrowth of the 
{\bf SCoLar} project by Bocharov and Bronstein. 
{\bf DELiA}, written in Turbo PASCAL by Bocharov (1991) and coworkers, 
is a commercial computer algebra system for investigating differential 
equations using Lie's approach. 

The program deals with symmetries, conservation laws, integrability and 
equivalence problems.
It has a special routine for systems of evolution equations, which we used
for computing conserved densities. 
We tested {\bf DELiA 1.5.1} on a PC with Intel 486 processor and 16 MB 
of RAM. 
Our test cases included equations (\ref{KdVeq}), (\ref{Boussys2}), 
(\ref{HirSatpar}),
(\ref{HirSatpar}) with $\alpha = 1/2$,  
(\ref{KdV5par}) with arbitrary $\alpha,$ $\beta = 20$ and $\gamma = 10\/$,
(\ref{Longwavesys}) and (\ref{3cKdV}). 

For (\ref{KdVeq}) and (\ref{HirSatpar}) with $\alpha = 1/2,$
{\bf DELiA} returned the same densities as the ones listed in this paper, 
up to terms that differ via integration by parts. 
With {\bf DELiA}, one cannot compute densities for (\ref{Boussys2}), 
(\ref{Longwavesys}), and (\ref{3cKdV}). These systems are out of its class:
the program requires second or higher-order spatial derivative terms in
all equations.
For systems with parameters, {\bf DELiA} does not automatically compute the 
densities corresponding to the (necessary) conditions on the parameters. 
One has to use {\bf DELiA}'s integrability test first, 
which determines conditions based on the existence of formal symmetries. 
Since these integrability conditions are neither necessary nor sufficient 
for the existence of conserved densities, one must further analyze the 
conditions manually.
Once the parameters are fixed, one can compute the densities. 
For further information we refer to Bocharov at alexei@wri.com.
\subsection{FS}

The REDUCE program {\bf FS} for ``formal symmetries'' was written 
by Gerdt and Zharkov (1990). 
The code {\bf FS} can be applied to polynomial nonlinear PDEs of 
evolution type, which are linear with respect to the highest-order 
spatial derivatives and with non-degenerated, diagonal coefficient matrix 
for the highest derivatives. The algorithm in {\bf FS} requires that
the evolution equation are of order two or higher in the
spatial variable. 
However, the formal symmetry approach does not require that the evolution 
equations are uniform in rank. 

We tested {\bf FS} with REDUCE 3.5 on an IBM Risc 6000 work station for 
equations (\ref{KdVeq}), (\ref{Boussys2}), (\ref{KdV5par}),
(\ref{Longwavesys}) and (\ref{3cKdV}).
We were unable to compute the densities for systems (\ref{Boussys2}), 
(\ref{Longwavesys}), and (\ref{3cKdV}), since {\bf FS} is not applicable
to such systems.
For (\ref{KdVeq}), {\bf FS} gave the same densities as our program,
up to terms that differ via integration by parts.

Like {\bf DELiA}, applied to equations with parameters, {\bf FS} computes
the conditions on the parameters using the symmetry approach. 
For (\ref{KdV5par}), {\bf FS} and {\bf condens.m} gave equivalent
sets of compatibility conditions on the parameters.
More information can be obtained from Gerdt at gerdt@jinr.dubna.su.
\subsection{Software under Development}

In addition to the available software, several research groups are
currently developing software for conserved densities.

Sanders and Wang at the Free University of Amsterdam are 
developing a software package in Maple and FORM to compute both conserved 
densities and recursion operators
(Sanders and Roelofs, 1994; Sanders and Wang, 1995a).
Their approach is more abstract and relies on the implicit function theorem. 
In contrast to our algorithm, their code makes no assumptions about the 
form of the conserved density.
Further information can be obtained from Sanders at jansa@cs.vu.nl. 

Wolf (1995) at the University of London is designing a package in REDUCE for 
the computation of conserved densities, which implements two methods.
If the density is of fixed order, then the program constructs it directly. 
This method is currently limited to two independent variables. 
In the second method the characteristic 
function of the conservation law is constructed first, and the density and 
flux are recovered via integration by parts. There is no limitation on 
the number of independent variables for this method.  
Both approaches use Wolf's program CRACK for solving overdetermined 
systems of PDEs.
See Hereman (1996) for a review of CRACK and for additional references 
to Wolf's work.
Wolf's algorithm is particularly efficient for showing the non-existence 
of conservation laws of high order. In contrast to our program, it also
allows one to compute non-polynomial conserved densities.
For further information contact Wolf at T.Wolf@maths.qmw.ac.uk.

Ahner, Tschantz, and Cook at Vanderbilt University are working on a 
similar project in {\it Mathematica} (Ahner, 1995).
We have no details about their algorithm or implementation. 
Contact Ahner at ahnerjf@ctrvax.vanderbilt.edu for further information.

Hickman (1996) at the University of Canterbury, Christchurch, New Zealand, 
has implemented a slight variation of our algorithm in Maple. Instead of 
computing the differential monomials in the density by repeated 
differentiation, 
Hickman uses a tree structure combining the appropriately weighted building 
blocks.
For further information, send email to M.Hickman@math.canterbury.ac.nz. 
\section{Conclusions}

A prototype of {\bf condens.m} was used to compute conserved densities of 
a generalized KdV equation due to Schamel (Verheest and Hereman, 1995), and 
a modified vector derivative NLS equation (Willox {\em et al.\/}, 1995).
Based on the results obtained with the software, integrability questions for 
these equations could be answered.

We offer the scientific community a {\it Mathematica} package to carry out 
the tedious calculations of conserved densities for systems of nonlinear 
evolution equations. 
Details about the algorithm and its implementation can be found in 
G\"{o}kta\c{s} (1996a), and G\"{o}kta\c{s} and Hereman (1997). 
The code {\bf condens.m}, together with several data and output files, 
is available via anonymous FTP (G\"{o}kta\c{s}, 1996b). 

Extensions of the algorithm presented in this paper towards PDEs with 
more than one spatial variable, dynamical systems, and systems of 
differential-difference equations are under development.
Further information about extensions can be obtained via email to 
ugoktas@mines.edu or whereman@mines.edu. 
\section*{Acknowledgements}
The authors are grateful to F. Verheest for many valuable comments and 
suggestions.
The Belgian National Fund for Scientific Research is thanked for a 
special research grant which made possible the stay of W.H. at the 
University of Ghent, where this work was started. 
\"{U}G appreciates the generous financial aid from the Turkish Government 
which supported his studies in part.
V. Gerdt, M. Hickman, M. Ito, J. Sanders and T. Wolf are thanked for
sharing information about their packages. 

\label{lastpage}
\end{document}